\documentclass[12pt,sort&compress]{elsarticle}

\usepackage{afterpage}
\usepackage{enumitem}
\usepackage{mystyle}
\usepackage{algpseudocode}
\usepackage{algorithm}
\usepackage{amsmath}
\usepackage{amssymb}
\usepackage{subfiles}
\graphicspath{{Images/}{../Images/}}
\usepackage{subfigmat} 
\usepackage{relsize}
\usepackage{pst-all}

\newcommand\figref[1]{Figure \ref{fig:#1}} 
\newcommand\tabref[1]{Table \ref{tab:#1}} 
\newcommand\eref[1]{Eq. (\ref{eq:#1})} 
\newcommand\p[1]{\partial{#1}}

\begin{document}

\begin{frontmatter}

\title{Invariant Data-Driven Subgrid Stress Modeling on Anisotropic Grids for Large Eddy Simulation}

\author[colorado]{Aviral Prakash \corref{cor1}}
\ead{aviral.prakash@colorado.edu}
\author[colorado]{Kenneth E. Jansen}
\author[colorado]{John~A.~Evans}
\cortext[cor1]{Corresponding author}

\address[colorado]{University of Colorado Boulder, Boulder, CO 80309, USA}
\begin{abstract}

We present a new approach for constructing data-driven subgrid stress models for large eddy simulation of turbulent flows using anisotropic grids.  The key to our approach is a Galilean, rotationally, reflectionally and unit invariant model form that also embeds filter anisotropy in such a way that an important subgrid stress identity is satisfied. We use this model form to train a data-driven subgrid stress model using only a small amount of anisotropically filtered DNS data and a simple and inexpensive neural network architecture. \textit{A priori} and \textit{a posteriori} tests indicate that the trained data-driven model generalizes well to filter anisotropy ratios, Reynolds numbers and flow physics outside the training dataset.
    
\end{abstract}


\end{frontmatter}

\section{Introduction}

Improvements in computational hardware have increasingly enabled scale-resolving simulations of complex turbulent flows. However, resolving all spatial and temporal scales with direct numerical simulation (DNS) is still computationally impractical for high Reynolds numbers flows. A viable alternative simulation methodology for many such flows is large eddy simulation (LES). In these simulations, larger turbulent structures that harbor most of the turbulent kinetic energy in the flow are resolved, while smaller isotropic turbulence scales, having a relatively lower turbulent kinetic energy content, are modeled. In LES, we solve the filtered Navier-Stokes equations that are unclosed due to the presence of the subgrid stress (SGS) tensor. The SGS tensor accounts for the interaction of unresolved scales with the resolved scales and accurate SGS models must be formulated to account for these interactions. Even though wall-resolved LES is computationally expensive for high Reynolds number industrial flows of interest \citep{Choi2012}, improvements in SGS models are also needed for accurate wall-modeled LES of complex flows \citep{Goc2021, Prakash2022b}.

Traditionally, SGS tensor models were developed based on physical observations and intuition \citep{Smagorinsky1963, Bardina1980} or mathematical simplifications to approximate the SGS tensor \citep{Clark1979, Stolz2001}. Several models like the Smagorinsky model \cite{Smagorinsky1963}, the WALE model \cite{Nicoud1999}, the $\sigma$-model \cite{Nicoud2011} and the $QR$ model \cite{Verstappen2011} involve a characteristic length scale which is selected to be the filter width. This filter width is often taken to be proportional to the local computational grid size. Even though applying these models to isotropic grids is straightforward, the length scale selection for anisotropic grids poses some issues. Several length scale formulations have been proposed over the years \citep{Deardorff1970, Scotti1992, Schumann1975}. However, these length scales are often insufficient for accounting for high grid anisotropy \citep{Haering2019, Trias2017}. Most turbulent flows often involve regions where using an anisotropic grid is essential for reducing the computational cost; for example, we require highly anisotropic grid resolution near the wall for wall-bounded flows. Therefore, developing SGS models that adapt well to grids with arbitrary anisotropy is essential. 

Recently, there have been considerable efforts towards developing data-driven closures for RANS and LES \cite{Duraisamy2019}. These models leverage high-fidelity DNS data and state-of-the-art machine learning-based regression techniques to construct a mapping between flow-based inputs and output closure terms. A comprehensive literature review of existing data-driven SGS models is presented in \citep{Prakash2021}. Even though there has been a large volume of work suggesting data-driven methodologies for developing SGS models, most strategies lead to models that do not conform to the physical symmetry or invariance properties of the SGS tensor. Some notable exceptions do embed physical invariance properties in the model construction. For example, several works \citep{Xie2020, Reissmann2021} employed the tensor-integrity basis approach \citep{Pope2000, Ling2016, Parmar2020} for constructing a rotationally and reflectionally invariant model form. Another method to build a Galilean, rotationally, reflectionally and unit invariant model form is shown in \cite{Prakash2021}. The data-driven SGS models proposed until now are either constructed for a fixed grid, that is, they either do not account for grid or flow-based length scale \citep{Gamahara2017, Wang2018} or have local computational grid stencil \citep{Maulik2017, Xie2020b}, or they use a scalar filter width as model input to characterize the computational grid \citep{Zhou2019, Xie2020, Reissmann2021, Prakash2021}. An arbitrary anisotropic grid cannot be defined entirely with a scalar characteristic length, so these models lose accuracy for anisotropic grids. These existing challenges motivate the need to develop data-driven SGS model formulations that adequately account for arbitrary filter anisotropy.

In this article, we propose an SGS model form that is not only Galilean, rotationally, reflectionally and unit invariant but also depends on filter anisotropy in such a manner that an important SGS anisotropy identity is satisfied.
We employ this model form to train a simple and inexpensive neural network SGS model using anisotropically filtered DNS data for forced HIT at $Re_{\lambda} = 418$. We conduct a series of \textit{a priori} and \textit{a posteriori} tests to demonstrate the accuracy of the trained model. For \textit{a priori} tests, we consider filter anisotropy outside the training dataset and observe the trained model yields accurate approximations of the exact SGS tensor. From \textit{a posteriori} tests, we observe that the trained model also generalizes well to cases involving filter anisotropy, Reynolds number and flow physics outside the training dataset, such as the anisotropic resolution of HIT at $Re_{\lambda} = \infty$  and the turbulent channel flow at $Re_{\tau} = 395$ and  $Re_{\tau} = 590$. 

An outline of this article is as follows. In Section \ref{section:FNSE}, we derive the filtered Navier-Stokes equations and introduce the SGS tensor that must be modeled in practice. In Section \ref{section:AnisotropicFilter}, we introduce the notion of an anisotropic filter kernel and derive a new SGS anisotropy identity used in constructing our new SGS model form. In Section \ref{section:ClassicalSGSModels}, we review some commonly used SGS tensor models and their application to anisotropic grids. In Section \ref{section:AnisotropicDDModel}, we provide details on constructing the proposed anisotropic model form. In Section \ref{section:TrainingModel}, we use the proposed model form to train a data-driven model. In Section \ref{section:Results}, we conduct \textit{a priori} and \textit{a posteriori} validation tests and compare the performance of the learned data-driven model against standard SGS models. In Section \ref{section:Conclusions}, we conclude by summarizing the model form development, highlighting key results and proposing directions for future research.

%

\section{The Filtered Navier-Stokes Equations}
\label{section:FNSE}

The incompressible Navier-Stokes equations are given as follows,

\begin{equation}
\frac{\partial u_i}{\p t} + \frac{\partial}{\p x_j} (u_i u_j)    = - \frac{1}{\rho} \frac{\partial p}{\p x_i} +\frac{\partial }{\p x_j} ( 2 \nu S_{ij} ) + f_i,
\end{equation}

\begin{equation}
\frac{\partial u_i}{\p x_i}    = 0,
\end{equation}

\noindent where $u_i$ is the $i^\text{th}$ component of the velocity field $\bm{u}$, $p$ is the pressure field, $\rho$ is the density, $\nu$ is the kinematic viscosity, $S_{ij} = \frac{1}{2}(\partial u_i/\partial x_j + \partial u_j/\partial x_i)$ is the ${ij}^\text{th}$ component of the strain-rate tensor and $f_i$ is the  $i^\text{th}$ component of the body force vector. 

In LES, larger resolved turbulent scales are separated from the smaller unresolved turbulent scales by a filtering operation. The filtering operation decomposes a flow variable $\phi$ as follows,

\begin{equation}
    \phi = \bar{\phi} + \phi',
\end{equation}

\noindent where $\bar{\phi}$ and $\phi^{'}$ are the filtered and sub-filter variables respectively. The filtering operation is mathematically defined as,

\begin{equation}
    \overline{\phi}(\bm{x}) = \int_{\mathbb{R}^3 } G(\bm{x},\bm{x}') \phi(\bm{x}') d^3 \bm{x}',
\end{equation}

\noindent where $G$ is known as the filter-kernel and $\mathbb{R}^3$ is the domain of filtering. Filtering is a linear operation that preserves constants, that is,

\begin{equation}
    \int_{\mathbb{R}^3 } a G(\bm{x},\bm{x}') d^3 \bm{x}' = a.
\end{equation}

\noindent A filter is known as a homogeneous filter if the filter kernel can be expressed as
\begin{equation}
    G(\bm{x},\bm{x}') = G_\text{homogeneous}( \bm{x} - \bm{x}').
\end{equation}

\noindent Furthermore, a filter is known as an isotropic filter if the filter kernel can be expressed as
\begin{equation}
    G(\bm{x},\bm{x}') = G_\text{isotropic}( |\bm{x} - \bm{x}'| ).
\end{equation}

We obtain the filtered Navier-Stokes equations by applying a homogeneous filter to the Navier-Stokes equations,

\begin{equation}
\frac{\partial \overline{u}_i}{\p t} + \frac{\partial}{\p x_j} (\overline{u}_i \overline{u}_j)    = - \frac{1}{\rho} \frac{\partial \overline{p}}{\p x_i} + \frac{\partial }{\p x_j} (2 \nu \overline{S}_{ij}) - \frac{\p \tau_{ij}}{\p x_j} + \overline{f}_i, \qquad \tau_{ij} = \overline{u_i u_j} - \bar{u}_i \bar{u}_j
\label{eq:FNS-momentum}
\end{equation}

\begin{equation}
    \frac{\partial \overline{u}_i}{\partial x_i} = 0, \label{eq:FNS-mass}
\end{equation}

\noindent where $\tau_{ij}$ is an unclosed term known as the subgrid stress (SGS) tensor. The SGS tensor is symmetric, Galilean invariant and unit invariant by definition.  Moreover, if the filter kernel is rotationally and reflectionally invariant, so is the SGS tensor.  Note that all isotropic filter kernels are rotationally and reflectionally invariant.  However, there are other filter kernels that are also rotationally and reflectionally invariant.

\section{Representation of Anisotropic Filters in a Parent Space}
\label{section:AnisotropicFilter}

 In this article, we focus on anisotropic filter kernels of the following form: 
 
\begin{equation}
    G(\pmb{x},\pmb{x}') = G_{\text{anisotropic}} ( \vert \pmb{A}^{-1} (\pmb{x} - \pmb{x}') \vert)
\end{equation}

\noindent where $\pmb{A}$ is a symmetric, positive definite tensor satisfying $\text{tr}(\pmb{A}^2) = \text{tr}(I) = 3$.  Note that such filter kernels are necessarily homogeneous, but they are not isotropic unless $\pmb{A} = \pmb{I}$.  Consequently, we refer to $\pmb{A}$ as the anisotropy tensor.  While anisotropic filter kernels are not necessarily isotropic, they are rotationally and reflectionally invariant and thus SGS tensors defined using an anisotropic filter kernel are also rotationally and reflectionally invariant.

As $\pmb{A}$ is a symmetric positive definite tensor, it admits the form:

\begin{equation}
    \pmb{A} = \lambda_1 \pmb{a}_1 \otimes \bm{a}_1 + \lambda_2 \pmb{a}_2 \otimes \bm{a}_2 + \lambda_3 \pmb{a}_3 \otimes \bm{a}_3.
\end{equation}

\noindent where $\lambda_i$ and $\pmb{a}_i$ are the $i^\text{th}$ eigenvalue and eigenvector respectively of $\pmb{A}$.  The eigenvectors correspond to the principal directions of filtering, while the eigenvalues give the ratio of the filter widths in each principal direction to an overall filter size.  In particular, if $\Delta_1$, $\Delta_2$ and $\Delta_3$ are the filter widths in directions $\pmb{a}_1$, $\pmb{a}_2$ and $\pmb{a}_3$, then $\lambda_1 = \Delta_1/\Delta$, $\lambda_2 = \Delta_2/\Delta$ and $\lambda_3 = \Delta_3/\Delta$ where $\Delta = \sqrt{(\Delta_1^2 + \Delta_2^2 + \Delta_3^2)/3}$.  Note that we can also construct a filter width tensor

\begin{equation}
    \bm{\Delta} = \Delta_1 \bm{a}_1 \otimes \bm{a}_1 +  \Delta_2 \bm{a}_2 \otimes \bm{a}_2 +  \Delta_3 \bm{a}_3 \otimes \bm{a}_3
\end{equation}

\noindent from the principal directions and filter widths in each principal direction and the filter width tensor is related to the anisotropy tensor through $\bm{A} = \bm{\Delta}/\Delta$.  Note while the filter width tensor is dimensional, the anisotropy tensor is non-dimensional.

A canonical example of an anisotropic filter kernel is the ellipsoidal filter kernel:

\begin{equation}
    G(\pmb{x},\pmb{x}') = G_{\text{anisotropic-ell}} ( \vert \pmb{A}^{-1} (\pmb{x} - \pmb{x}') \vert) = \left\{ 
\begin{array}{cc}
\frac{6}{\pi \Delta_1 \Delta_2 \Delta_3} & \text{ if } \vert \pmb{A}^{-1} (\pmb{x} - \pmb{x}') \vert < \frac{\Delta}{2} \\
0 & \text{ otherwise. }
\end{array}
 \right.
\end{equation}

\begin{figure}[t!]
    \def\svgwidth{0.4\textwidth}
    \centering
    \includegraphics[width=0.6\textwidth]{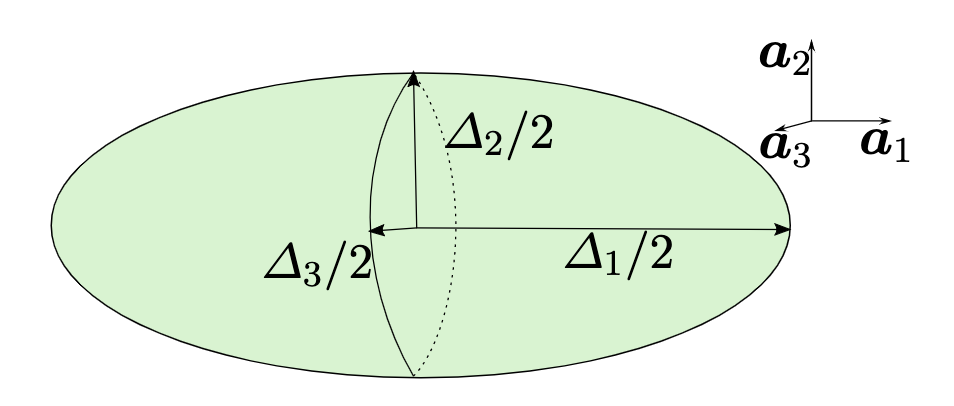}
    \caption{The filter domain associated with an ellipsoidal filter kernel.}
    \label{fig:AnisotropicFilterDom}
\end{figure}

\noindent The filter domain for this kernel, defined as the region for which the kernel is nonzero, is an ellipsoid centered at $\bm{x}$ whose semi-axes are oriented in directions $\bm{a}_1$, $\bm{a}_2$ and $\bm{a}_3$ and have lengths $\Delta_1/2$, $\Delta_2/2$ and $\Delta_3/2$ respectively.  This is graphically depicted in \figref{AnisotropicFilterDom}.

Let us now define a linear mapping $\pmb{\xi}: \mathbb{R}^3 \to \mathbb{R}^3$ as follows:

\begin{equation}
    \pmb{\xi}(\pmb{x}) = \pmb{A}^{-1} \pmb{x}.
    \label{eq:pullback}
\end{equation}

\begin{figure}[b!]
    \centering
    \def\svgwidth{\textwidth}
    \includegraphics[width=\textwidth]{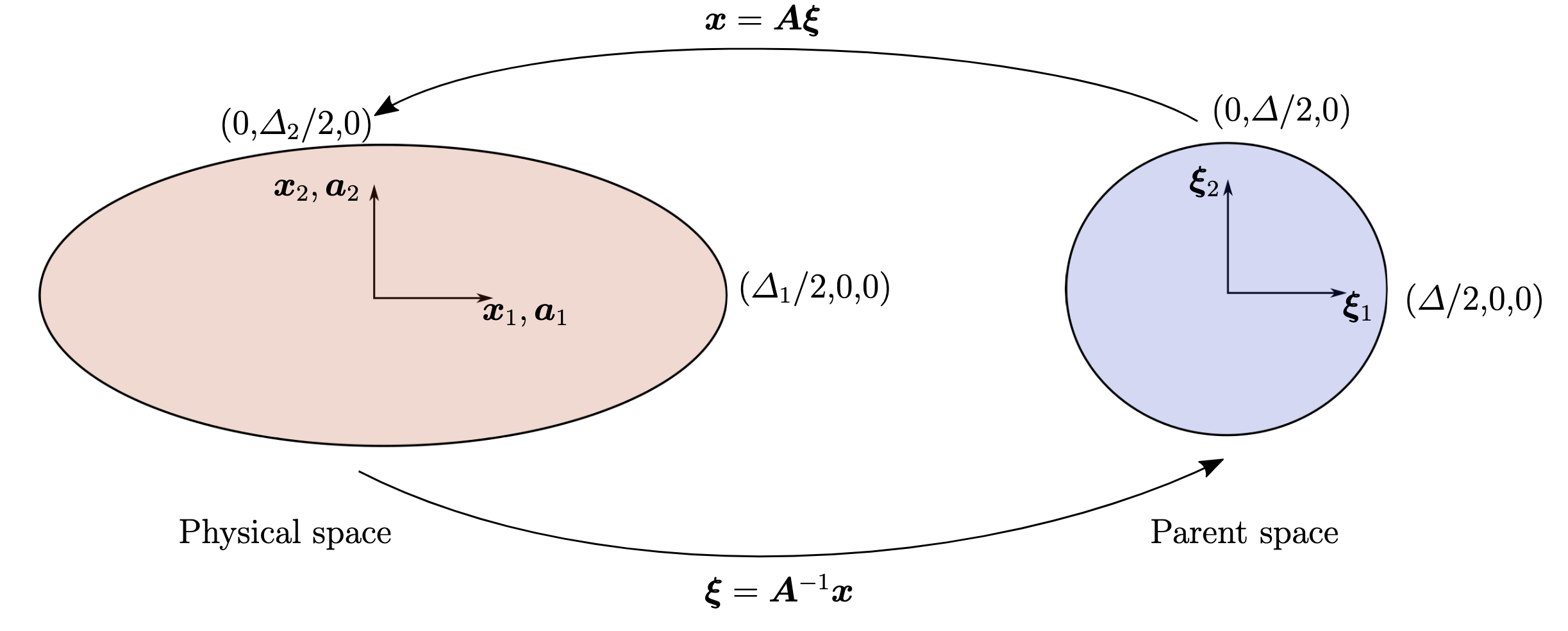}    
    \caption{Physical space and parent space.}
    \label{fig:AnisoGrid}
\end{figure}

\begin{figure}[t!]
    \def\svgwidth{\textwidth}
    \includegraphics[width=\textwidth]{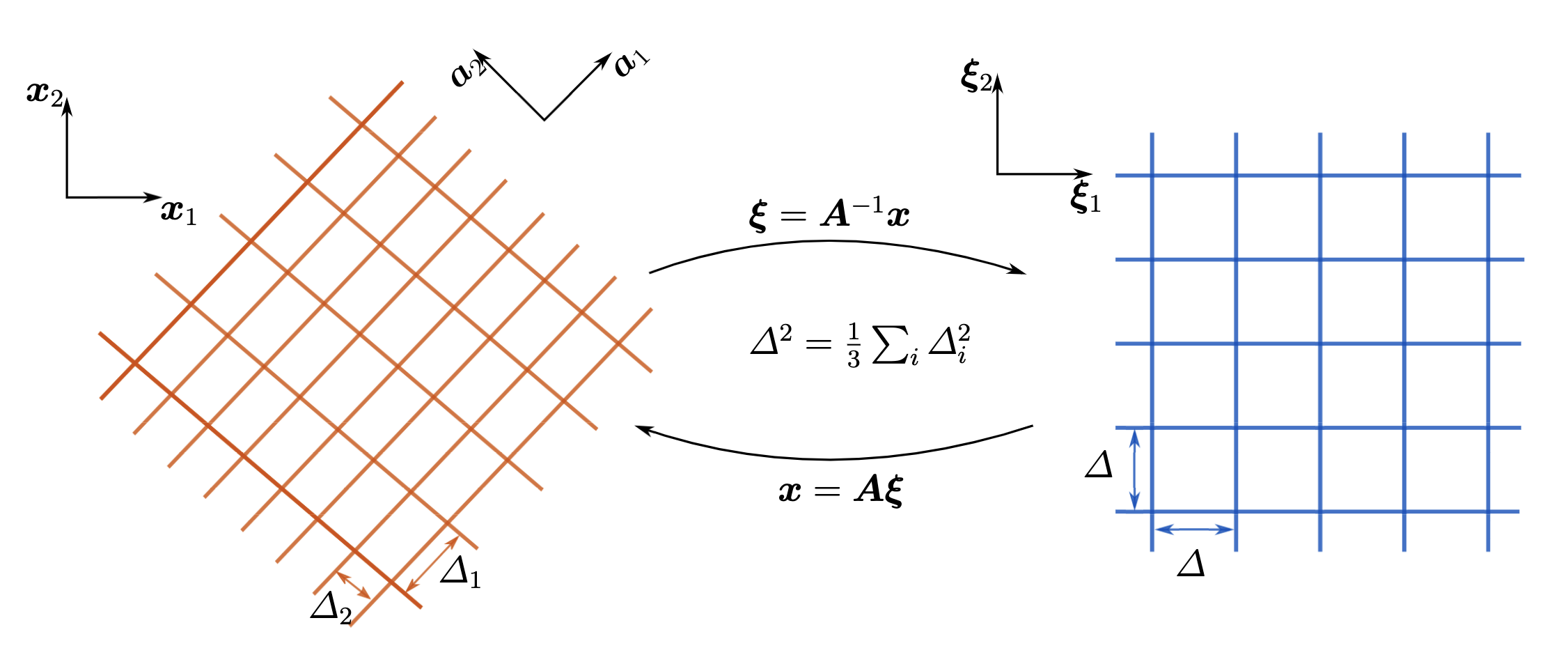}       
    \caption{Mapping an anisotropic grid in physical space to an isotropic grid in parent space.}
    \label{fig:AnisoGridBox}
\end{figure}

\noindent We refer to the domain and range of the above mapping as the physical space and parent space, respectively and we refer to coordinates in the physical space as physical coordinates and coordinates in the parent space as parent coordinates.  As depicted in \figref{AnisoGrid}, the linear mapping $\pmb{\xi}: \mathbb{R}^3 \to \mathbb{R}^3$ maps ellipsoids in the physical space with semi-axis directions $\bm{a}_1$, $\bm{a}_2$ and $\bm{a}_3$ and lengths $\Delta_1/2$, $\Delta_2/2$ and $\Delta_3/2$ to spheres in the parent space with radii $\Delta/2$ and it also has an inverse

\begin{equation} 
    \pmb{x}(\pmb{\xi}) = \pmb{A} \pmb{\xi}
    \label{eq:pushforw}
\end{equation}

\noindent that maps spheres in the parent space to ellipsoids in the physical space.  The linear mapping $\pmb{\xi}: \mathbb{R}^3 \to \mathbb{R}^3$ also maps anisotropic grids in the physical space with grid sizes $\Delta_1$, $\Delta_2$ and $\Delta_3$ in directions $\bm{a}_1$, $\bm{a}_2$ and $\bm{a}_3$ to an isotropic grid in the parent space with grid size $\Delta$ as depicted in \figref{AnisoGridBox}.

Since

\begin{equation}
    \int_{\R^3} G_{\text{anisotropic}} ( \vert \pmb{A}^{-1} (\pmb{x} - \pmb{x}') \vert) d \pmb{x}' = \int_{\R^3} G ( \pmb{x}, \pmb{x}') d \pmb{x}' \nonumber = 1
\end{equation}

\noindent we can change variables to parent coordinates to arrive at:

\begin{equation}
    \int_{\R^3} G_{\text{anisotropic}} ( \vert \pmb{\xi} - \pmb{\xi}' \vert) \text{det} (\pmb{A}) d \pmb{\xi}' = 1.
\end{equation}

\noindent Defining

\begin{equation}
    G_{\text{isotropic}}( r) := 
    G_{\text{anisotropic}} ( r) \text{det} (\pmb{A}), \label{eq:mapped_filter}
\end{equation}

\noindent it follows that

\begin{equation}
    \int_{\R^3} G_{\text{isotropic}} ( \vert \pmb{\xi} - \pmb{\xi}' \vert) d \pmb{\xi}' = 1.
\end{equation}

\noindent Consequently,
\begin{equation}
G_{\text{parent}}(\pmb{\xi}, \pmb{\xi}') := G_{\text{isotropic}} ( \vert \pmb{\xi} - \pmb{\xi}' \vert)
\end{equation}
is a suitable parent space filter kernel.  Moreover, it is an isotropic filter kernel, as opposed to the physical space filter kernel $G(\bm{x},\bm{x}')$.  However, the parent space filter kernel $G_{\text{parent}}(\pmb{\xi}, \pmb{\xi}')$ and physical space filter kernel $G(\bm{x},\bm{x}')$ are connected through \eqref{eq:mapped_filter}. Unsurprisingly, we can also connect physical space filtered quantities (which we henceforth denote using ($\bar{\cdot}$)) to parent space filtered quantities (which we henceforth denote using ($\tilde{\cdot}$)).  To see this, let $\phi$ be a field defined over physical space.  Then by a change of variables

\begin{align}
    \bar{\phi} ( \bm{x}) &= \int_{\R^3} \phi ( \bm{x}' ) G_{\text{anisotropic}}(\vert \pmb{A}^{-1}(\bm{x} - \bm{x}') \vert) d\bm{x}' \nonumber \\ &= \int_{\R^3} \phi (\pmb{A} \bm{\xi}' ) G_{\text{isotropic}}(\vert \bm{\xi} - \bm{\xi}' \vert) d\bm{\xi}' \nonumber \\ &= (\widetilde{\phi \circ \bm{x}}) ( \bm{\xi (\bm{x}}) ).
\end{align}

\noindent By an identical calculation, we have

\begin{equation}
    \bar{\bm{u}} (\bm{x}) = (\widetilde{\bm{u} \circ \bm{x}}) ( \bm{\xi}(\bm{x}) ).
    \label{eq:parent_vel}
\end{equation}

\noindent The above indicates that the physical space filtered velocity field is precisely the parent space filtered velocity field (more precisely, the pushforward to the physical space of the filter of the pullback to the parent space of the velocity field).  By yet another identical calculation, we have

\begin{equation}
    \overline{u_i u_j} (\bm{x}) = \widetilde{(u_i \circ \bm{x}) (u_j \circ \bm{x})} ( \bm{\xi}(\bm{x}) ),
    \label{eq:parent_velproduct}
\end{equation}

\noindent and it follows that

\begin{equation}
    \overline{u_i u_j} (\bm{x}) - \bar{u}_i (\bm{x}) \bar{u}_j (\bm{x}) = \widetilde{(u_i \circ \bm{x}) (u_j \circ \bm{x})} ( \bm{\xi}(\bm{x}) ) - (\widetilde{u_i \circ \bm{x}}) ( \bm{\xi}(\bm{x}) ) \; (\widetilde{u_j \circ \bm{x}}) ( \bm{\xi}(\bm{x}) ). \label{eq:anisotropy_identity}
\end{equation}

\noindent The above equation relates the physical space SGS tensor to an analogous tensor in the parent space that we henceforth refer to as the parent space SGS tensor.  We refer to \eqref{eq:anisotropy_identity} as the SGS tensor anisotropy identity and we will use it later to embed anisotropy into a novel SGS model form.

\section{Classical SGS Models}
\label{section:ClassicalSGSModels}

Classical SGS models are commonly based on physical approximations of turbulent flow behavior, for example, alignment of the deviatoric part of the SGS stress tensor and the resolved strain-rate tensor \cite{Smagorinsky1963} \cite{Germano1991} or similarity between the smallest resolved and largest unresolved scales \cite{Bardina1980}. One of the most commonly used SGS models, the Smagorinsky model \cite{Smagorinsky1963}, approximates the deviatoric part of the SGS tensor, denoted as $\tau_{ij}^d$, using the equation: 

\begin{equation}
    \tau^{d, Smag}_{ij} = -2 \nu_t \bar{S}_{ij}, \qquad \nu_t = L (\bm{\Delta})^2 \vert \bar{\pmb{S}} \vert,
\end{equation}

\noindent where $L(\bm{\Delta}) = C_s \Delta_e$ is associated turbulence length scale \cite{Scotti1992}. This length scale is often taken proportional to an effective filter width $\Delta_e$ and this proportionality constant is known as the Smagorinsky constant. Several different definitions of $\Delta_e$ have been considered in the literature. Among these definitions are: the geometric mean of filter width components, $\Delta_e = (\Delta_1 \Delta_2 \Delta_3)^{1/3}$, maximum value of filter width component $\Delta_e =  \text{max}\{\Delta_1, \Delta_2, \Delta_3\}$ and average of the norm of filter width components $\Delta_e = \Big(1/3(\Delta_1^2 + \Delta_2^2 + \Delta_3^2) \Big)^{1/2}$. Identically, these definitions correspond to the $(\text{det} (\pmb{\Delta}))^{1/3}$, $ || \pmb{\Delta} ||_2$ and $|| \pmb{\Delta} ||_F/\sqrt{3} $ respectively. These definitions work well for mild anisotropies. However, they fail to accurately represent strong filter anisotropy effects \citep{Scotti1992}. By considering energy transfer equilibrium between resolved and unresolved turbulent scales for isotropic turbulence, a scaling for the geometric mean definition is proposed in \citep{Scotti1992}:

\begin{equation}
    \Delta_e = (\Delta_1 \Delta_2 \Delta_3)^{1/3} f(a_1, a_2), \quad f(a_1, a_2) = \text{cosh} \sqrt{\frac{4}{27}\Big[ (\ln a_1)^2 - \ln a_1  \ln a_2 + (\ln a_2)^2 \Big]},
\end{equation}

\noindent where $a_1$ and $a_2$ are aspect ratios of two smaller filter width components to the largest filter width component. However, this definition is inadequate for pencil-type filters \citep{Haering2019}. The dynamic procedure was proposed in \citep{Germano1991, Germano1992} to determine the optimal value of the Smagorinsky constant for accurate turbulent flow statistics. This procedure involves the application of explicit test filtering of the flow field to obtain the value ($C_S L(\bm{\Delta}))^2$. In the resulting model, often known as the dynamic Smagorinsky model \cite{Germano1991} \cite{Germano1992},  the length scale is dynamically determined, which overcomes the selection of an effective filter width definition. The analysis of energy spectra for the Smagorinsky model with different filter width specifications and dynamically determined length scale showed that all versions of the model exhibited inadequate representation for scales smaller than wavenumber corresponding to the grid cutoff of largest resolved direction \cite{Scotti1997}. The dynamic Smagorinsky model also involves averaging the model constant over homogeneous directions due to stability constraints. This averaging procedure could lead to loss of anisotropy sensitivity for the smaller scales \cite{Haering2019}. Comparison of the performance of the Smagorinsky model with the geometric mean length scale, anisotropic minimum dissipation (AMD) model \cite{Rozema2015} and M43 model in the presence of resolution anisotropy for forced homogeneous isotropic turbulence (HIT) test case was performed in \cite{Haering2019}. They observed that the AMD model and M43 model give a much better prediction of the energy spectra at higher wavenumbers in the direction of resolution anisotropy than the Smagorinsky model. An inertial tensor based on the local grid element geometry has also been used to account for grid anisotropy \citep{Abba2017}. The components of the inertial tensor can be represented as a $\Delta_{ik} \Delta_{jk}$ up to a scaling factor. In addition to these filter-based length scales, flow-based length scales are often used in SGS modeling, however, we will not discuss them in this article. Interested readers should refer to work by \citep{Piomelli2015} for more details on flow-based length scale formulations. The characteristics and applicability of several filter width and flow-based length scale approximations are also well summarised in \cite{Schumann2020} \cite{Trias2017}. Even though several length scale definitions have been proposed over the years, there is no common consensus on the optimal definition \cite{Haering2019}. 

Other SGS models based on mathematical approximations of the filtering operation have also been proposed over the years. One of the most common ones amongst them is the gradient model \cite{Clark1979}, which is often characterized as a part of a bigger class of approximate deconvolution models (ADM) \cite{Stolz2001}. As these models are based on mathematical approximations such as a Taylor series expansion in Fourier space \citep{Clark1979} or approximate deconvolution based on Van-Cittert iteration \citep{Stolz2001}, filter anisotropy is inherently considered in the model form. For example, the gradient model in anisotropic form can be represented as, 

\begin{equation}
    \tau^{GM}_{ij} = \frac{1}{12} \bar{G}_{im} \Delta_{mk} \bar{G}_{jn} \Delta_{nk}
    \label{eq:GM_eq}
\end{equation}

\noindent where $\bar{G}_{ij} = \partial \bar{u}_i/ \partial x_j$ is the velocity gradient tensor. If the axis of filtering is aligned to the coordinate axis, the filter width tensor reduces to,

\begin{equation}
    \begin{bmatrix}\Delta_{ij}\end{bmatrix} = \begin{bmatrix}
    \Delta_1 & 0 & 0 \\
    0 & \Delta_2 & 0 \\
    0 & 0 & \Delta_3 \\
    \end{bmatrix},
\end{equation}

\noindent resulting in the following anisotropic form of the gradient model:

\begin{equation}
    \tau^{GM}_{ij} = \frac{1}{12} \Big[ \Delta_1^2 \frac{\partial \bar{u}_i}{\partial x_1} \frac{\partial \bar{u}_j}{\partial x_1} + \Delta_2^2 \frac{\partial \bar{u}_i}{\partial x_2} \frac{\partial \bar{u}_j}{\partial x_2} + \Delta_3^2 \frac{\partial \bar{u}_i}{\partial x_3} \frac{\partial \bar{u}_j}{\partial x_3} \Big].
\end{equation}

\noindent This anisotropic form of the gradient model is used in \cite{Trias2017} to derive a flow-based length scale approximation for anisotropic filters. Even though the anisotropic form of the gradient model is well suited for anisotropic filters, it has not been as popular as eddy viscosity models, such as the dynamic Smagorinsky model, possibly due to under-dissipative nature of the gradient model leading to energy pileup at resolved scales for high Reynolds number flows. In this article, we will use the dynamic Smagorinsky model and the anisotropic form of the gradient model for comparison against the data-driven model that we train using anisotropically filtered DNS data.

\section{Data-Driven Modeling of the SGS Tensor for Anisotropic Filters}
\label{section:AnisotropicDDModel}

\subsection{Existing Data-Driven Modeling Techniques}

Several data-driven SGS modeling techniques have been suggested over recent years. As data-driven SGS modeling is an active research field, it is impossible to cover all data-driven closure techniques exhaustively. One way to broadly classify data-driven SGS models is through the length scale specification. Many data-driven models \citep{Maulik2017, Xie2020b} express the stress tensor in terms of non-local flow variables, that is, the SGS tensor depends on flow variables at the stencil of surrounding points. Most often, as the grid stencil is fixed, these models do not explicitly depend on filter width. Without a specification of filter width in the input space, the extension of these approaches to significantly larger filter widths outside the training dataset is questionable. Other data-driven approaches \citep{Zhou2019, Xie2020, Reissmann2021, Prakash2021} include a scalar filter width in the input space. These models can exhibit good prediction of stresses for filter widths that are outside the training dataset. In particular, the data-driven model proposed in \cite{Prakash2021} satisfies physical invariance properties and exhibits good generalization properties for filter widths, Reynolds number and flow physics outside the training dataset in both \textit{a priori} and \textit{a posteriori} tests. Since these data-driven models take in a scalar filter width for turbulent length scale specification, they can be used in the presence of anisotropic grids by using equivalent filter width definitions mentioned in  Section \ref{section:ClassicalSGSModels}. However, it has been well documented that these length scale definitions do not adequately represent an arbitrary filter anisotropy \citep{Schumann2020, Haering2019}. In this article, we propose a solution to this problem by introducing a data-driven SGS model that extends to arbitrary anisotropic filters while ensuring physical invariance properties.

\subsection{Construction of Anisotropic Model Form}

We inspire the selection of model inputs and outputs from work done in \cite{Prakash2021} to satisfy Galilean invariance. For their isotropic model, the following form is selected:

\begin{equation}
    \tau = \tau^{\text{\text{model}}} (\bar{\bm{G}}, \bm{I}, \Delta) = \tau^{\text{\text{model}}} (\bar{\bm{G}}, \Delta).
\end{equation}

\noindent where $\bm{I}$ is the identity tensor that indicates the isotropy of the filter. In addition to these model inputs, the SGS tensor in anisotropic physical space also depends on the orientation of the filter quantified by the anisotropy tensor. Therefore, the model form becomes:

\begin{equation}
\label{eq:original_form}
    \tau_{ij} = \overline{u_i u_j} - \bar{u}_i \bar{u}_j = \tau^{\text{\text{model}}} (\bar{\bm{G}}, \bm{A}, \Delta)
\end{equation}

\noindent where the inputs can be composed of any combinations of $\bar{\bm{G}}$, $\bm{A}$ and $\Delta$. If the model form satisfies the SGS tensor anisotropy identity, we can express the SGS tensor in terms of flow variables in the mapped space. In this case, we have the alternative model form,

\begin{equation}
    \tau_{ij} = \widetilde{(u_i \circ \bm{x}) (u_j \circ \bm{x})} ( \bm{\xi}(\bm{x}) ) - (\widetilde{u_i \circ \bm{x}}) ( \bm{\xi}(\bm{x}) ) \; (\widetilde{u_j \circ \bm{x}}) ( \bm{\xi}(\bm{x}) ) = \tau^{\text{\text{model}}} (\tilde{\bm{G}}, \bm{I}, \Delta), \label{eq:embedmodelform}
\end{equation}

\noindent where $\tilde{G}_{ij} = \partial (\widetilde{u_i \circ x})/ \partial \xi_j$ is the gradient of filtered velocity in the parent filter space. This quantity is related to the velocity gradient in physical anisotropic space as follows,

\begin{equation}
    \tilde{G}_{ij} = \frac{\partial (\widetilde{u_i \circ x})}{\partial \xi_j} = \frac{\partial (\widetilde{u_i \circ x})}{\partial x_k} \frac{\partial x_k}{\partial \xi_j} = \frac{\partial \bar{u}_i}{\partial x_k} \frac{\partial x_k}{\partial \xi_j} = \bar{G}_{ik} A_{kj}.
\end{equation}

\noindent The model form given by \eqref{eq:embedmodelform} ensures that the constructed SGS model exactly satisfies the SGS tensor anisotropy identity and thereby embeds filter anisotropy.  Moreover, any model of the form given by \eqref{eq:original_form} that satisfies the SGS tensor anisotropy identity must also be of the form given by \eqref{eq:embedmodelform}, so no generality is lost.

As the SGS tensor anisotropy identity provides us with an expression for the SGS tensor in terms of velocity in the parent filter space, we construct SGS models following the same strategy as the one used for constructing SGS models for isotropic filters as shown in \cite{Prakash2021}. We consider the symmetric and anti-symmetric part of the gradient of filtered velocity in the parent filter space,

\begin{equation}
    \tilde{S}_{ij} = \tilde{G}_{ij}^{\;\text{sym}} = \frac{\tilde{G}_{ij} + \tilde{G}_{ji}}{2},
\end{equation}

\begin{equation}
    \tilde{\Omega}_{ij} = \tilde{G}_{ij}^{\;\text{anti-sym}} = \frac{\tilde{G}_{ij} - \tilde{G}_{ji}}{2},
\end{equation}

\noindent and refer to them as $\tilde{S}$ and $\tilde{\Omega}$ tensor. For an isotropic grid, these tensors reduce to standard filtered strain-rate and rotation-rate tensors. With these inputs, the resulting model is given as,
\begin{equation}
\label{eq:finalanisomodel_form}
    \tau = \tau^{\text{\text{model}}} (\bm{\tilde{S}}, \bm{\tilde{\Omega}}, \Delta).
\end{equation}
It is clear that this model form is analogous to the one defined used for isotropic models in \citep{Prakash2021} with the difference in the definitions of $\tilde{\bm{S}}$ and $\tilde{\bm{\Omega}}$. In addition to this set of inputs, we also consider kinematic viscosity in the set of inputs. As briefly discussed in \cite{Prakash2021}, the data-driven model with an added viscosity input gave better predictions in the transition region for the Taylor-Green Vortex case at $Re = 1600$. Furthermore, in the past, the Van-Driest damping function \cite{Vandriest1956} involving a viscous length scale has been used to improve the near-wall behavior of the Smagorinsky model. Based on numerical experiments, we observed that adding viscosity as an input allows the model to have a superior near-wall behavior by appropriately turning off the model in over-resolved flow regions, typically near the wall. The final dimensional SGS tensor model form is given as follows,
\begin{equation}
\label{eq:finalanisomodel_form_2}
    \tau = \tau^{\text{\text{model}}} (\bm{\tilde{S}}, \bm{\tilde{\Omega}}, \Delta, \nu).
\end{equation}

Next, we represent the model inputs and outputs in a form that satisfies rotational and reflectional invariance, thereby ensuring that the SGS model form is invariant to these transformations. In particular, we consider the approach proposed in \cite{Prakash2021} involving the representation of model inputs and outputs in the coordinate frame corresponding to the eigen-frame of the filtered strain-rate tensor. In this article, we extend this approach to anisotropic grids by choosing the eigen-frame for the symmetric part of the gradient of filtered velocity in the parent filter space as the coordinate frame for representing our inputs and outputs. We refer to this flow-based coordinate frame as the $\tilde{S}$-frame. The components of the $\tilde{S}$ tensor in the $\tilde{S}$-frame comprise a diagonal matrix,

\begin{equation}
    \begin{bmatrix}
    \tilde{\mathrm{S}}_{ij}^{\tilde{S}}
    \end{bmatrix} = 
    \begin{bmatrix}
    \lambda_1^{\Tilde{S}} & 0 & 0 \\
    0 & \lambda_2^{\tilde{S}} & 0 \\
    0 & 0 & \lambda_3^{\tilde{S}} 
    \end{bmatrix},
\end{equation}

\noindent where $\lambda_1^{\Tilde{S}}$, $\lambda_2^{\Tilde{S}}$ and $\lambda_3^{\Tilde{S}}$ are the eigenvalues of $\tilde{S}$ tensor. The eigenvalues are ordered as follows,

\begin{equation}
     \lambda^{\tilde{S}}_1 \geq \lambda^{\tilde{S}}_2 \geq \lambda^{\tilde{S}}_3.
\end{equation}

\noindent The components of the $\tilde{\Omega}$ tensor in the $\tilde{S}$-frame comprise an antisymmetric matrix,

\begin{equation}
    \begin{bmatrix}
    \tilde{\mathrm{\Omega}}_{ij}^{\tilde{S}}
    \end{bmatrix} = \frac{1}{2}
    \begin{bmatrix}
    0 & \tilde{\omega}_3^{\Tilde{S}} & -\tilde{\omega}_2^{\Tilde{S}} \\
    -\tilde{\omega}_3^{\Tilde{S}} & 0 & \tilde{\omega}_1^{\Tilde{S}} \\
    \tilde{\omega}_2^{\Tilde{S}} & -\tilde{\omega}_3^{\Tilde{S}} & 0 
    \end{bmatrix},
\end{equation}

\noindent where $\tilde{\omega}_1^{\Tilde{S}}$, $\tilde{\omega}_2^{\Tilde{S}}$ and $\tilde{\omega}_3^{\Tilde{S}}$ are the elements of $\tilde{\Omega}$ in the $\tilde{S}$-frame. The model output, that is the SGS tensor, is represented in the $\tilde{S}$-frame as follows,

\begin{equation}
    \tau_{ij}^{\tilde{S}} = V^{\tilde{S}}_{ki} \; \tau_{kl} \; V^{\tilde{S}}_{lj},
\end{equation}

\noindent where $V^{\tilde{S}}_{ij}$ is the $i^\text{th}$ component of the $j^\text{th}$ eigenvector of the $\tilde{S}$ tensor. The selection and orientation of eigenvectors follow a strategy similar to the one suggested in \cite{Prakash2021} with the key difference being the use of $\tilde{\bm{\omega}}$ instead of vorticity to align the eigenvectors. The components of the final model form are as follows,

\begin{equation}
\tau_{ij}^{\tilde{S}} = \tau_{ij}^{\tilde{S},\text{model}} ( \lambda_1^{\tilde{S}}, \lambda_2^{\tilde{S}}, \lambda_3^{\tilde{S}}, \tilde{\omega}_1^{\tilde{S}}, \tilde{\omega}_2^{\tilde{S}}, \tilde{\omega}_3^{\tilde{S}}, \Delta, \nu).
\end{equation}

\noindent In the original model form, \eref{original_form}, we have as model inputs the nine components of the velocity gradient tensor, the six unique components of the anisotropy tensor and one filter width component, therefore a total of sixteen inputs. By representing the components of anisotropic filtered velocity gradient in the eigen-frame of $\tilde{S}$ tensor, we have reduced the number of inputs to only eight. Furthermore, this input set is of a considerably smaller size than the minimal tensor integrity basis often used for RANS \cite{Ling2016} \cite{Wang2018} \cite{Peters2020} and LES \cite{Xie2020} \cite{Reissmann2021} closure modeling. In what follows, we also replace the input $\tilde{\omega}_3^{\tilde{S}}$ by the following input,

\begin{equation}
\begin{split}
    \tilde{G} & = \Big( \tilde{S}_{ij} \tilde{S}_{ij} + \tilde{\Omega}_{ij} \tilde{\Omega}_{ij} \Big)^{1/2} = \Big( \tilde{S}^{\tilde{S}}_{ij} \tilde{S}^{\tilde{S}}_{ij} + \tilde{\Omega}^{\tilde{S}}_{ij} \tilde{\Omega}^{\tilde{S}}_{ij} \Big)^{1/2} \\
    & = \Bigg( (\lambda_1^{\tilde{S}})^2 + (\lambda_2^{\tilde{S}})^2 + (\lambda_3^{\tilde{S}})^2 + \frac{1}{2}\Big( (\omega_1^{\tilde{S}})^2 + (\omega_2^{\tilde{S}})^2 + (\omega_3^{\tilde{S}})^2 \Big) \Bigg)^{1/2},
\end{split}    
\end{equation}

\noindent yielding the alternative but equivalent model form,

\begin{equation}
\tau_{ij}^{\tilde{S}} = \tau_{ij}^{\tilde{S},\text{model}} ( \lambda_1^{\tilde{S}}, \lambda_2^{\tilde{S}}, \lambda_3^{\tilde{S}}, \tilde{\omega}_1^{\tilde{S}}, \tilde{\omega}_2^{\tilde{S}}, \tilde{G}, \Delta, \nu).
\end{equation}

Lastly, we incorporate unit invariance in the model form by using the Buckingham-Pi theorem. As we have nine terms and two independent physical units (length and time), we must have seven $\Pi$ variables. The resulting model is of the form,

\begin{equation}
\Pi_7 = \hat{\tau}_{ij}^{\tilde{S},\text{model}} (\Pi_1, \Pi_2, \Pi_3, \Pi_4, \Pi_5, \Pi_6).
\end{equation}

\noindent We define the $\Pi$ variables as follows,

\begin{equation}
    \Pi_1 = \hat{\lambda}_1^{\tilde{S}} = \frac{\lambda_1^{\tilde{S}}}{\tilde{G}},
\end{equation}

\begin{equation}
    \Pi_2 = \hat{\lambda}_2^{\tilde{S}}  = \frac{\lambda_2^{\tilde{S}}}{\tilde{G}},
\end{equation}

\begin{equation}
    \Pi_3 = \hat{\lambda}_3^{\tilde{S}}  = \frac{\lambda_3^{\tilde{S}}}{\tilde{G}},
\end{equation}

\begin{equation}
    \Pi_4 = \hat{\tilde{\omega}}_1^{\tilde{S}} = \frac{\tilde{\omega}_1^{\tilde{S}}}{\tilde{G}},
\end{equation}

\begin{equation}
    \Pi_5 = \hat{\tilde{\omega}}_2^{\tilde{S}} = \frac{\tilde{\omega}_2^{\tilde{S}}}{\tilde{G}},
\end{equation}

\begin{equation}
    \Pi_6 = \hat{\nu} = \frac{\nu}{\Delta^2 \tilde{G}},
\end{equation}

\begin{equation}
    \Pi_7 = \hat{\tau}^{\tilde{S}}_{ij} = \frac{\tau_{ij}^{\tilde{S}}}{\Delta^2 \tilde{G}^2},
\end{equation}

\noindent yielding the final model form,

\begin{equation}
\tau_{ij}^{\tilde{S}} = \Delta^2 \tilde{G}^2 \hat{\tau}_{ij}^{\tilde{S},\text{model}} (\hat{\lambda}_1^{\tilde{S}}, \hat{\lambda}_2^{\tilde{S}}, \hat{\lambda}_3^{\tilde{S}}, \hat{\omega}_1^{\tilde{S}}, \hat{\omega}_2^{\tilde{S}}, \hat{\nu}). \label{eq:final_model_form}
\end{equation}

\noindent The final non-dimensional model form for the SGS tensor, given in \eref{final_model_form}, satisfies: 1) Galilean, 2) rotational, 3) reflectional, and 4) unit invariance and also embeds filter anisotropy in such a way that the SGS anistropy identity is satisfied.

\subsection{Model Form Representation of the Gradient Model}

The anisotropic form of the gradient model can be obtained from the proposed anisotropic model form. To show this, we first look at the derivation of the anisotropic form of the gradient model. For a box-filter kernel, we take the Fourier transform of the SGS tensor representation in the parent filter space, perform a Taylor series expansion, truncate the higher-order terms and take an inverse Fourier transform of the resulting expansion to obtain the expression of the gradient model in the parent filter space,

\begin{equation}
    \tau_{ij} = \frac{\Delta^2}{12} \tilde{G}_{ik} \tilde{G}_{jk}.
    \label{eq:grad_model}
\end{equation}

\noindent By substituting $\tilde{G}_{ij} = G_{ik} A_{kj}$, we obtain, 

\begin{equation}
    \tau_{ij} = \frac{\Delta^2}{12} G_{ik} A_{kl} G_{jm} A_{ml} = \frac{1}{12} G_{ik} \Delta_{kl} G_{jm} \Delta_{ml},
\end{equation}

\noindent which is the same as the anisotropic form of the gradient model, \eref{GM_eq}. Note that expressing \eref{grad_model} in the $\tilde{S}$-frame results in the equation
\begin{equation}
    \tau^{\tilde{S}}_{ij} = \frac{\Delta^2}{12} \tilde{G}_{ik}^{\tilde{S}} \tilde{G}_{jk}^{\tilde{S}}
\end{equation}
\noindent where
\begin{equation}
    \begin{bmatrix}
    \mathrm{\tilde{G}}_{ij}^{\tilde{S}}
    \end{bmatrix} = 
    \begin{bmatrix}
    \mathrm{\tilde{S}}_{ij}^{\tilde{S}}
    \end{bmatrix} + \begin{bmatrix}
    \mathrm{\tilde{\Omega}}_{ij}^{\tilde{S}}
    \end{bmatrix} = 
    \begin{bmatrix}
    \lambda_1^{\tilde{S}} & \omega_3^{\tilde{S}} /2 & -\omega_2^{\tilde{S}} /2 \\
    -\omega_3^{\tilde{S}} /2 & \lambda_2^{\tilde{S}} & \omega_1^{\tilde{S}} /2 \\
    \omega_2^{\tilde{S}} /2 & -\omega_1^{\tilde{S}} /2 & \lambda_3^{\tilde{S}}
    \end{bmatrix}.
\end{equation}
Thus can express the SGS tensor predicted by the gradient model in the $\tilde{S}$-frame as follows:

\begin{equation}
\tau_{ij}^{\tilde{S}} = \Delta^2 \tilde{G}^2 \hat{\tau}_{ij}^{\tilde{S},\text{gradient}} (\hat{\lambda}_1^{\tilde{S}}, \hat{\lambda}_2^{\tilde{S}}, \hat{\lambda}_3^{\tilde{S}}, \hat{\omega}_1^{\tilde{S}}, \hat{\omega}_2^{\tilde{S}}). 
\label{eq:final_model_form_V2}
\end{equation}

\noindent where

\begin{equation}
    \hat{\tau}_{11}^{\tilde{S},\text{gradient}}(\hat{\lambda}_1^{\tilde{S}},\hat{\lambda}_2^{\tilde{S}},\hat{\lambda}_3^{\tilde{S}},\hat{\omega}_1^{\tilde{S}},\hat{\omega}_2^{\tilde{S}}) = (\hat{\lambda}_1^{\tilde{S}})^2 + \frac{1}{4} (\hat{\omega}_2^{\tilde{S}})^2  + \frac{1}{4} (\hat{\omega}_3^{\tilde{S}})^2
\end{equation}

\begin{equation}
    \hat{\tau}_{22}^{\tilde{S},\text{gradient}}(\hat{\lambda}_1^{\tilde{S}},\hat{\lambda}_2^{\tilde{S}},\hat{\lambda}_3^{\tilde{S}},\hat{\omega}_1^{\tilde{S}},\hat{\omega}_2^{\tilde{S}}) = (\hat{\lambda}_2^{\tilde{S}})^2 + \frac{1}{4} (\hat{\omega}_1^{\tilde{S}})^2  + \frac{1}{4} (\hat{\omega}_3^{\tilde{S}})^2
\end{equation}

\begin{equation}
    \hat{\tau}_{33}^{\tilde{S},\text{gradient}}(\hat{\lambda}_1^{\tilde{S}},\hat{\lambda}_2^{\tilde{S}},\hat{\lambda}_3^{\tilde{S}},\hat{\omega}_1^{\tilde{S}},\hat{\omega}_2^{\tilde{S}}) = (\hat{\lambda}_3^{\tilde{S}})^2 + \frac{1}{4} (\hat{\omega}_1^{\tilde{S}})^2  + \frac{1}{4} (\hat{\omega}_2^{\tilde{S}})^2
\end{equation}

\begin{equation}
    \hat{\tau}_{12}^{\tilde{S},\text{gradient}}(\hat{\lambda}_1^{\tilde{S}},\hat{\lambda}_2^{\tilde{S}},\hat{\lambda}_3^{\tilde{S}},\hat{\omega}_1^{\tilde{S}},\hat{\omega}_2^{\tilde{S}}) = \hat{\tau}_{21}^{\tilde{S},\text{gradient}}(\hat{\lambda}_1^{\tilde{S}},\hat{\lambda}_2^{\tilde{S}},\hat{\lambda}_3^{\tilde{S}},\hat{\omega}_1^{\tilde{S}},\hat{\omega}_2^{\tilde{S}}) = \frac{1}{2} (\hat{\lambda}_1^{\tilde{S}} - \hat{\lambda}_2^{\tilde{S}}) \hat{\omega}_3^{\tilde{S}} - \frac{1}{4} \hat{\omega}_1^{\tilde{S}} \hat{\omega}_2^{\tilde{S}}
\end{equation}

\begin{equation}
    \hat{\tau}_{13}^{\tilde{S},\text{gradient}}(\hat{\lambda}_1^{\tilde{S}},\hat{\lambda}_2^{\tilde{S}},\hat{\lambda}_3^{\tilde{S}},\hat{\omega}_1^{\tilde{S}},\hat{\omega}_2^{\tilde{S}}) = \hat{\tau}_{31}^{\tilde{S},\text{gradient}}(\hat{\lambda}_1^{\tilde{S}},\hat{\lambda}_2^{\tilde{S}},\hat{\lambda}_3^{\tilde{S}},\hat{\omega}_1^{\tilde{S}},\hat{\omega}_2^{\tilde{S}}) = \frac{1}{2} (\hat{\lambda}_3^{\tilde{S}} - \hat{\lambda}_1^{\tilde{S}}) \hat{\omega}_2^{\tilde{S}} - \frac{1}{4} \hat{\omega}_1^{\tilde{S}} \hat{\omega}_3^{\tilde{S}}
\end{equation}

\begin{equation}
    \hat{\tau}_{23}^{\tilde{S},\text{gradient}}(\hat{\lambda}_1^{\tilde{S}},\hat{\lambda}_2^{\tilde{S}},\hat{\lambda}_3^{\tilde{S}},\hat{\omega}_1^{\tilde{S}},\hat{\omega}_2^{\tilde{S}}) = \hat{\tau}_{32}^{\tilde{S},\text{gradient}}(\hat{\lambda}_1^{\tilde{S}},\hat{\lambda}_2^{\tilde{S}},\hat{\lambda}_3^{\tilde{S}},\hat{\omega}_1^{\tilde{S}},\hat{\omega}_2^{\tilde{S}}) = \frac{1}{2} (\hat{\lambda}_2^{\tilde{S}} - \hat{\lambda}_3^{\tilde{S}}) \hat{\omega}_1^{\tilde{S}} - \frac{1}{4} \hat{\omega}_2^{\tilde{S}} \hat{\omega}_3^{\tilde{S}}
\end{equation}

\noindent and

\begin{equation}
    \hat{\omega}_3^{\tilde{S}} = \Big( 2 - 2 (\hat{\lambda}_1^{\tilde{S}})^2 + 2 (\hat{\lambda}_2^{\tilde{S}})^2 + 2 (\hat{\lambda}_3^{\tilde{S}})^2 - (\hat{\omega}_1^{\tilde{S}})^2 - (\hat{\omega}_2^{\tilde{S}})^2 \Big)^{1/2}.
\end{equation}

\noindent This expression of the gradient model is a quadratic polynomial in terms of model inputs. Therefore, the gradient model can be written similarly to the proposed non-dimensional model form without dependence on $\hat{\nu}$. In other words, the proposed anisotropic model form allows for a more generalized expression for the SGS model with the same model inputs as the anisotropic form of the gradient model.

\subsection{Functional Mapping Using Artificial Neural Networks}

The next step is to learn a functional mapping between inputs and outputs using regression techniques. In this article, we use artificial neural networks (ANNs) for this purpose. By increasing the number of neurons in each layer or the number of layers, we are capable of representing increasingly nonlinear mappings. For a cost-effective SGS model, we select a neural network architecture with a single layer and 20 neurons to represent the model. The training procedure for ANNs utilizes an optimization algorithm (stochastic gradient descent algorithm) to obtain the optimal values of weights and biases that minimize the specified cost functional. In this article, we utilize ANNs for learning a non-linear functional mapping between the six inputs (each of $\hat{\lambda}_1^{\tilde{S}}, \hat{\lambda}_2^{\tilde{S}}, \hat{\lambda}_3^{\tilde{S}}, \hat{\omega}_1^{\tilde{S}}, \hat{\omega}_2^{\tilde{S}}, \hat{\nu}$) and six outputs (each component of $\hat{\tau}^{\tilde{S}}_{ij}$). A comprehensive description of ANNs is out of the scope of this paper, however, interested readers could refer to \cite{Goodfellow2016} for more details.

\section{A Simple Anisotropic Data-Driven Model for the SGS Tensor}
\label{section:TrainingModel}

A Galilean, rotationally, reflectionally and unit invariant SGS model for anisotropic filters can be learned by using the model form derived in the previous section and training the model using anisotropically filtered DNS data. We extract raw DNS data for forced homogeneous and isotropic turbulence (HIT) flow at $Re_{\lambda} = 418$ from the Johns Hopkins Turbulence Database (JHTDB) \cite{Li2008}. We further apply anisotropic filters consisting of several filter widths on DNS data, as shown in \tabref{Training} and obtain the training/testing dataset. The smallest filter width is the same size as the grid resolution of DNS data; therefore, the SGS stress tensor is zero for that filter width. Adding this smallest filter width to the training dataset adds information as to when the learned SGS model should turn off. The rest of the filter widths have the same aspect ratio corresponding to a pencil-type filter ($AR = \frac{\Delta_3}{\Delta_1} = \frac{\Delta_3}{\Delta_2} > 1$) at increasing base filter widths ($\Delta_1$). Only a small amount of data is used for training the model. In particular, other than the smallest filter width, only a single aspect ratio with a pencil-type anisotropic filter is used for the training data. We hypothesize that by embedding physical invariance properties and filter anisotropy in the model form, ANNs require only a limited amount of data to generalize well to scenarios outside the training dataset.

\begin{table}[ht!]
    \centering
    \resizebox{\columnwidth}{!}{%
    \begin{tabular}{ccccc}
        \hline
        \hline
         \textbf{Dataset} & \textbf{No. of samples} & \textbf{Spatial Locations} & \textbf{Time} & \textbf{Filter Width} \\
        &&&& $\Delta_1 \; \text{x} \;\Delta_2 \;\text{x} \;\Delta_3$ \\
         \hline
         Training/Testing & $196,608$/$65,536$ & Randomly sampled in  & t = 1s & 2.2 \text{x} 2.2 \text{x} 2.2 $\eta$ \\
          &  &  $ x_i \in [0.5\pi ,\; 1.5\pi]$ &  & 2.2 \text{x} 2.2 \text{x} 6.6 $\eta$ \\
          &  &  &  & 6.6 \text{x} 6.6 \text{x} 19.8 $\eta$ \\
          &  &  &  & 15.4 \text{x} 15.4 \text{x} 46.2 $\eta$ \\
          &  &  &  & 28.6 \text{x} 28.6 \text{x} 85.8 $\eta$ \\
          &  &  &  & 46.2 \text{x} 46.2 \text{x} 138.6 $\eta$ \\
         \hline
         \hline
    \end{tabular}
    }
    \caption{Training dataset}
    \label{tab:Training}
\end{table}

In this article, we select the mean squared error between the modeled and exact non-dimensionalized filtered SGS stresses in the $\tilde{S}$-frame as the loss function for optimizing the weights and biases of ANNs. The loss function is given as, 

\begin{equation}
    \text{MSE} (\boldsymbol{\hat{W}},\boldsymbol{\hat{b}}) = \frac{1}{n_\text{train}} \sum_{a = 1}^{n_\text{train}} \sum_{i = 1}^{3} \sum_{j = 1}^{3} \left( \hat{\tau}_{ij}^{\tilde{S},\textup{DNS}}(\textbf{x}_a) - \hat{\tau}_{ij}^{\tilde{S},\textup{model}} \left(\hat{\boldsymbol{q}}^\text{DNS}(\textbf{x}_a); \boldsymbol{\hat{W}}, \boldsymbol{\hat{b}}\right) \right)^2 \;,
    \label{eq:MSE}
\end{equation}

\noindent where $\hat{\tau}_{ij}^{\tilde{S},\textup{model}}$ denotes the ANN model for the $ij^\text{th}$ component of the non-dimensional SGS tensor in the $S$-frame, $\boldsymbol{\hat{W}}$ and $\boldsymbol{\hat{b}}$ denote the weights and biases of the ANN model, $\hat{\tau}_{ij}^{\tilde{S},\textup{DNS}}$ denotes the DNS value of the $ij^\text{th}$ component of the non-dimensional SGS tensor in the $\tilde{S}$-frame, $ \hat{\boldsymbol{q}}^{\textup{DNS}}$ denotes the DNS value of the non-dimensional input vector (composed of $\hat{\lambda}_1^{\tilde{S}}$, $\hat{\lambda}_2^{\tilde{S}}$, $\hat{\lambda}_3^{\tilde{S}}$, $\hat{\omega}_1^{\tilde{S}}$, $\hat{\omega}_2^{\tilde{S}}$ and $\hat{\nu}$) and $\left\{ \textbf{x}_a \right\}_{a=1}^{n_\text{train}}$ denotes the set of training points. To assess the training convergence, we also evaluate the correlation coefficient between the non-dimensionalized modeled SGS tensor ($\hat{\tau}_{ij}^{\tilde{S},\text{model}}$) and the exact SGS tensor ($\hat{\tau}_{ij}^{\tilde{S},\text{DNS}}$) in $\tilde{S}$-frame:

\begin{equation}
    \text{C.C.}^{\tilde{S}} = \mathlarger{\sum_{i}} \mathlarger{\sum_{j}} \frac{\langle ( \hat{\tau}^{\tilde{S},\text{DNS}}_{ij}- \langle \hat{\tau}^{\tilde{S},\text{DNS}}_{ij} \rangle) ( \hat{\tau}^{\tilde{S},\text{\text{model}}}_{ij} - \langle \hat{\tau}^{\tilde{S},\text{\text{model}}}_{ij} \rangle) \rangle}{(\langle (\hat{\tau}^{\tilde{S},\text{DNS}}_{ij}- \langle \hat{\tau}^{\tilde{S},\text{DNS}}_{ij} \rangle)^2  \rangle)^{1/2}(\langle (\hat{\tau}^{\tilde{S},\text{\text{model}}}_{ij}- \langle \hat{\tau}^{\tilde{S},\text{\text{model}}}_{ij} \rangle)^2  \rangle)^{1/2}}.
    \label{eq:CCs}
\end{equation}

\begin{figure}[t!]
    \centering
    \subfigure[\label{fig:Training_MSE}]{\includegraphics[width=0.49\textwidth]{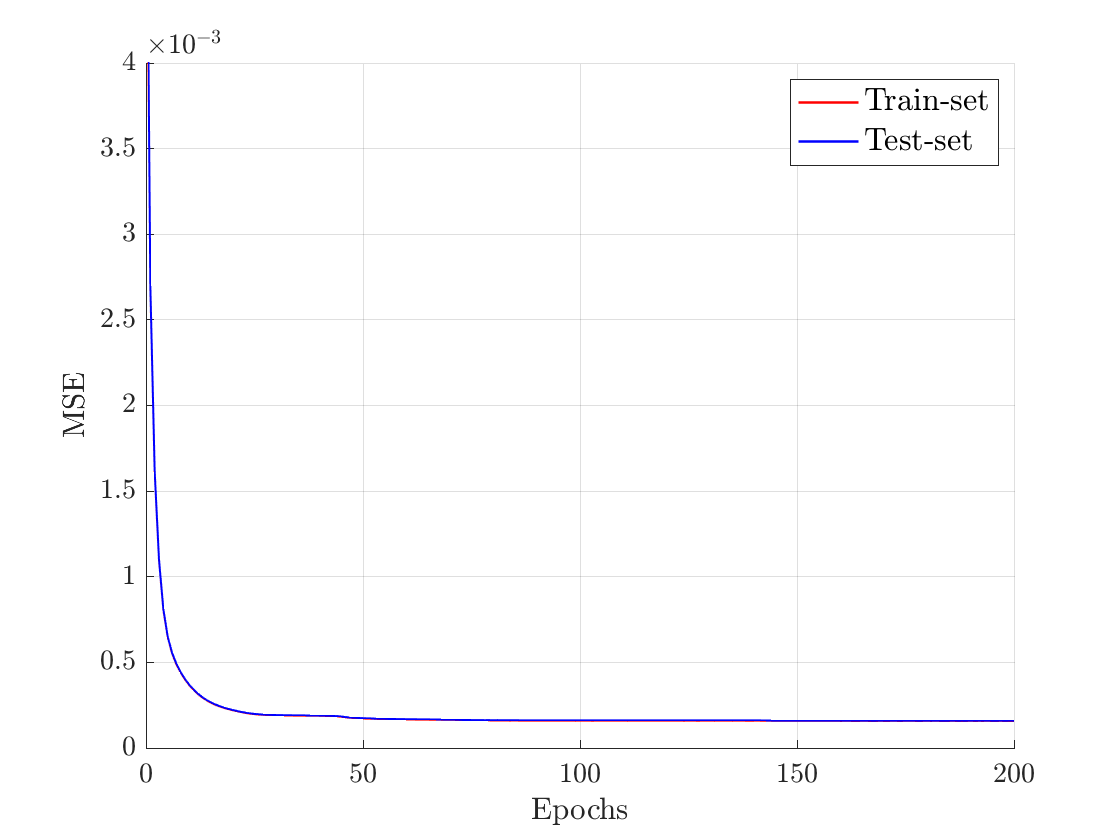}}
    \subfigure[\label{fig:Training_CCs}]{\includegraphics[width=0.49\textwidth]{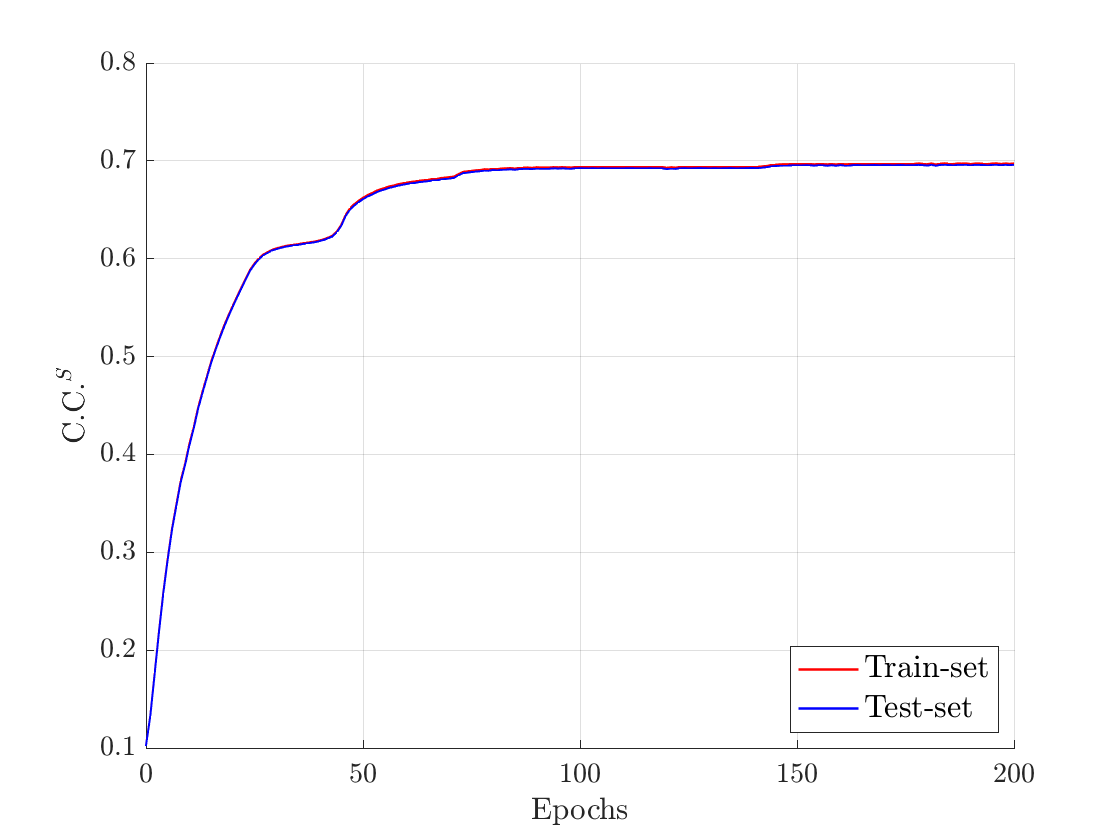}}
    \caption{Convergence characteristics for the ANN model training: (a) mean-squared error ($\text{MSE}$) and (b) non-dimensional $S$-frame correlation coefficient ($\text{C.C.}^S$).}
    \label{fig:Training}
\end{figure}

\noindent The convergence of MSE and $\text{C.C.}^{\tilde{S}}$ are shown in \figref{Training}. We observe the asymptotic behavior of these two quantities at higher epochs indicating sufficient model convergence. 

\section{Numerical Results}
\label{section:Results}

The learned data-driven model is evaluated using \textit{a priori} and \textit{a posteriori} tests and compared to classical SGS models. \textit{A priori} tests involve comparing the modeled SGS tensor to the exact SGS tensor extracted by filtering DNS data. These tests are quicker to evaluate and give an initial estimate of the model performance. On the other hand, \textit{a posteriori} tests involve performing an LES. Even though these tests are expensive to conduct, they are more comprehensive and enable assessment of both model accuracy and stability for different flows. The list of SGS models used in this article and their corresponding abbreviations are summarized in \tabref{abbrev_SGS}

\begin{table}[t!]
    \centering
    \begin{tabular}{cc}
        \hline
        \hline
         \textbf{\text{Model}} & \textbf{Abbreviation}  \\
         \hline
         No Model & NM \\
         Dynamic Smagorinsky Model & DSM \\
         Gradient Model & GM \\
         Data-driven Model & DD \\
         \hline
         \hline
    \end{tabular}
    \caption{List of SGS models compared in this article.}
    \label{tab:abbrev_SGS}
\end{table}

\subsection{\textit{A priori} results}

We first perform \textit{a priori} tests by comparing the modeled SGS stress with those obtained by filtering data from the JHTDB. The tests are performed for pencil-type and book-type anisotropic filtered data at several aspect ratios that are not a part of the training set (shown in \tabref{Validation_pencil} and \tabref{Validation_book}).
\begin{table}[b!]
    \centering
    \resizebox{\columnwidth}{!}{%
    \begin{tabular}{ccccc}
        \hline
        \hline
         \textbf{Dataset} & \textbf{No. of samples} & \textbf{Spatial Locations} & \textbf{Time} & \textbf{Aspect Ratio} \\
        &&&& $\Delta_1 \; \text{x} \;\Delta_2 \;\text{x} \;\Delta_3$ \\
         \hline
         Validation & $262,144$ & Randomly sampled in  & t = 1s & 15.4 \text{x} 15.4 \text{x} 15.4 $\eta$ \\
          &  & $ x_i \in [0.5\pi ,\; 1.5\pi]$ &  & 15.4 \text{x} 15.4 \text{x} 77 $\eta$ \\
          &  &  &  & 15.4 \text{x} 15.4 \text{x} 169.4 $\eta$ \\
          &  &  &  & 15.4 \text{x} 15.4 \text{x} 231 $\eta$ \\
          &  &  &  & 15.4 \text{x} 15.4 \text{x} 292.6 $\eta$ \\
         \hline
         \hline
    \end{tabular}
    }
    \caption{Validation dataset for pencil-type anisotropic filters}
    \label{tab:Validation_pencil}
\end{table}
\begin{table}[ht!]
    \centering
    \resizebox{\columnwidth}{!}{%
    \begin{tabular}{ccccc}
        \hline
        \hline
         \textbf{Dataset} & \textbf{No. of samples} & \textbf{Spatial Locations} & \textbf{Time} & \textbf{Aspect Ratio} \\
        &&&& $\Delta_1 \; \text{x} \;\Delta_2 \;\text{x} \;\Delta_3$ \\
         \hline
         Validation & $262,144$ & Randomly sampled in  & t = 1s & 15.4 \text{x} 15.4 \text{x} 15.4 $\eta$ \\
          &  &  $ x_i \in [0.5\pi ,\; 1.5\pi]$ &  & 15.4 \text{x} 77 \text{x} 77 $\eta$ \\
          &  &  &  & 15.4 \text{x} 169.4 \text{x} 169.4 $\eta$ \\
          &  &  &  & 15.4 \text{x} 231 \text{x} 231 $\eta$ \\
          &  &  &  & 15.4 \text{x} 292.6 \text{x} 292.6 $\eta$ \\
         \hline
         \hline
    \end{tabular}
    }
    \caption{Validation dataset for book-type anisotropic filters}
    \label{tab:Validation_book}
\end{table}
The model performance can be categorized by two quantities: correlation coefficient (C.C.) and relative error in mean energy flux. The correlation coefficient between modeled and exact SGS stresses is defined as

\begin{equation}
    \text{C.C.} = \mathlarger{\sum_{i}} \mathlarger{\sum_{j}} \frac{\langle ( \tau^{\textup{DNS}}_{ij}- \langle \tau^{\textup{DNS}}_{ij} \rangle) ( \tau^{M}_{ij} - \langle \tau^{M}_{ij} \rangle) \rangle}{(\langle (\tau_{ij}^{\textup{DNS}}- \langle \tau_{ij}^{\textup{DNS}} \rangle)^2  \rangle)^{1/2}(\langle (\tau^{M}_{ij}- \langle \tau^{M}_{ij} \rangle)^2  \rangle)^{1/2}}
    \label{eq:CC},
\end{equation}

\noindent where $\tau^{M}_{ij}$ and $\tau^{\textup{DNS}}_{ij}$ are the $ij^{th}$ components of the modeled and exact SGS tensors. This estimate is often used to gauge the structural accuracy of a model, that is values closer to $1$ correspond to a more structurally accurate SGS model. We also define relative error in mean energy flux as,

\begin{equation}
    \text{R.E.F.} = \frac{\langle \Pi^M \rangle - \langle \Pi^\text{DNS} \rangle }{\langle \Pi^\text{DNS} \rangle}
    \label{eq:REF},
\end{equation}

\noindent where $\Pi_M = - \tau^M_{ij} S_{ij}$ and $\Pi_{DNS} = - \tau^{\textup{DNS}}_{ij} S_{ij}$ are the modeled and exact SGS dissipation respectively. This quantity identifies the dissipative performance of the model. A positive value indicates over-dissipation; conversely, a negative value points to under-dissipation for the given case. These two quantities together serve as a good preliminary examination for SGS models. 

\begin{figure}
    \centering
    \subfigure[\label{fig:CC_Apriori_Book_7dx}]{\includegraphics[width=0.49\textwidth]{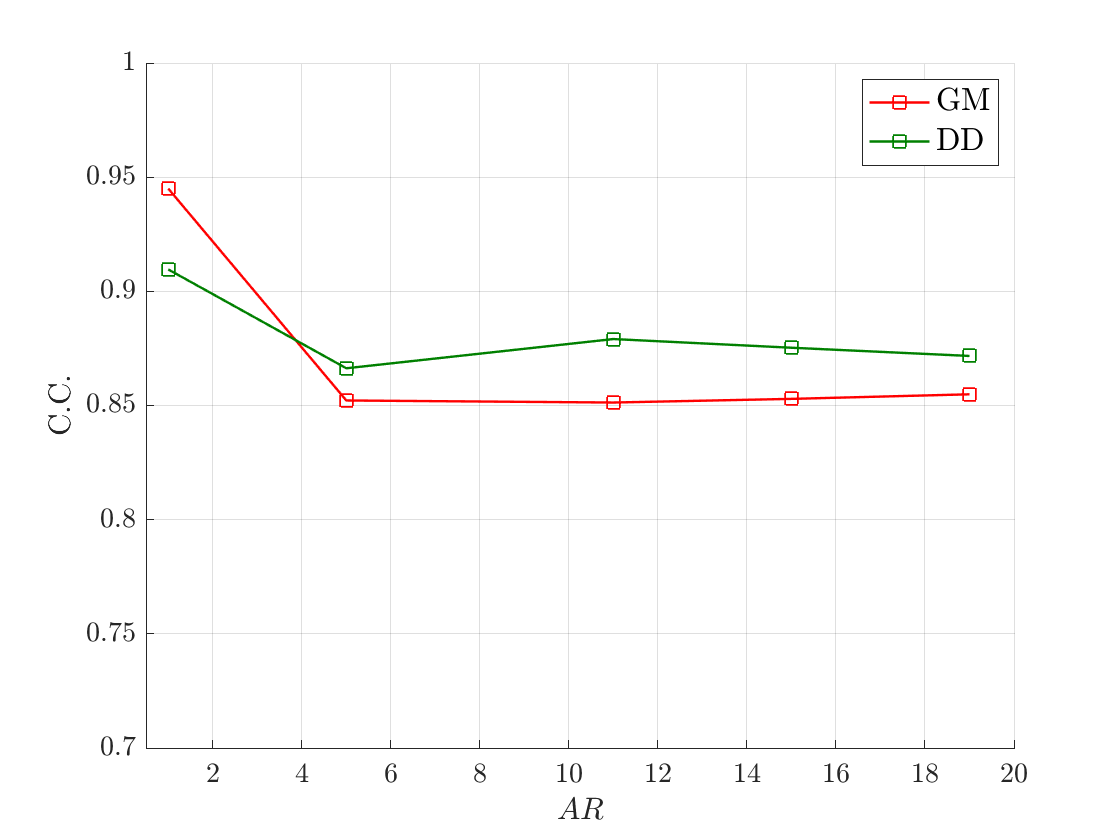}}
    \subfigure[\label{fig:REF_Apriori_Book_7dx}]{\includegraphics[width=0.49\textwidth]{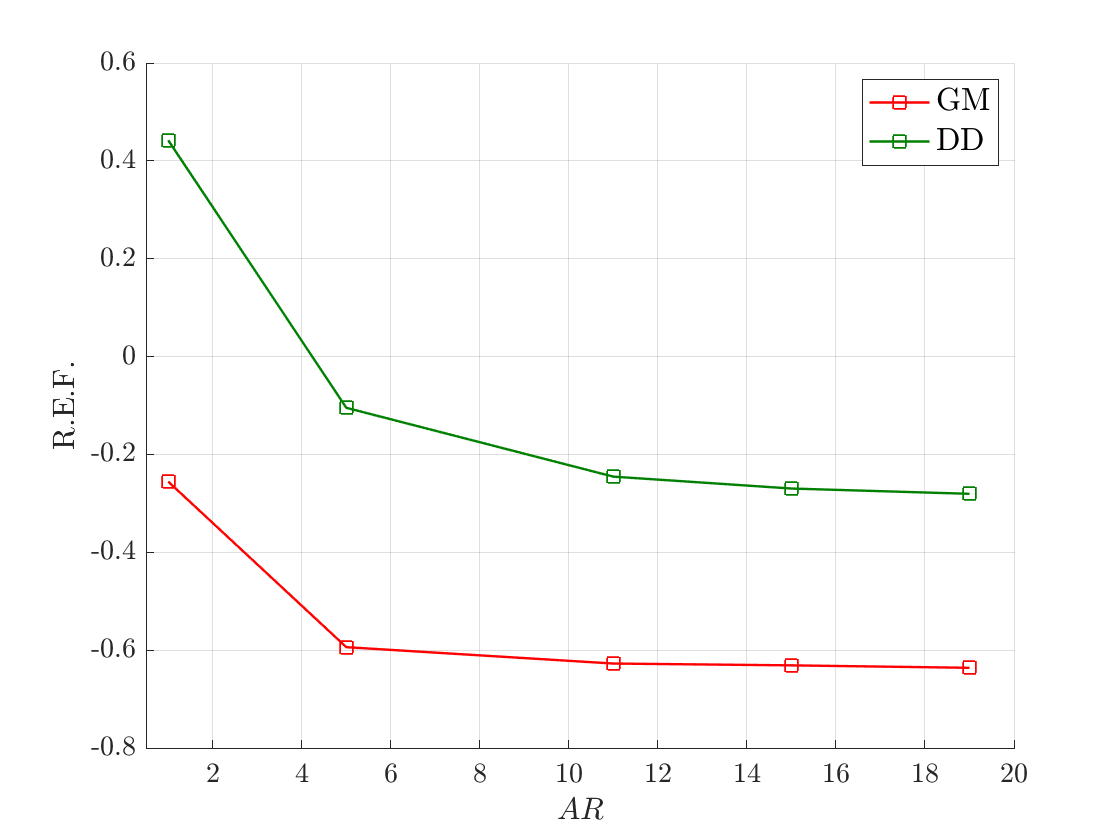}}
    \caption{(a) Correlation Coefficient (C.C.) and (b) Relative error in mean energy flux (R.E.F) for book-type anisotropic filters}
    \label{fig:Apriori_Book_7dx}
\end{figure}

\begin{figure}[t]
    \centering
    \subfigure[\label{fig:CC_Apriori_Pencil_7dx}]{\includegraphics[width=0.49\textwidth]{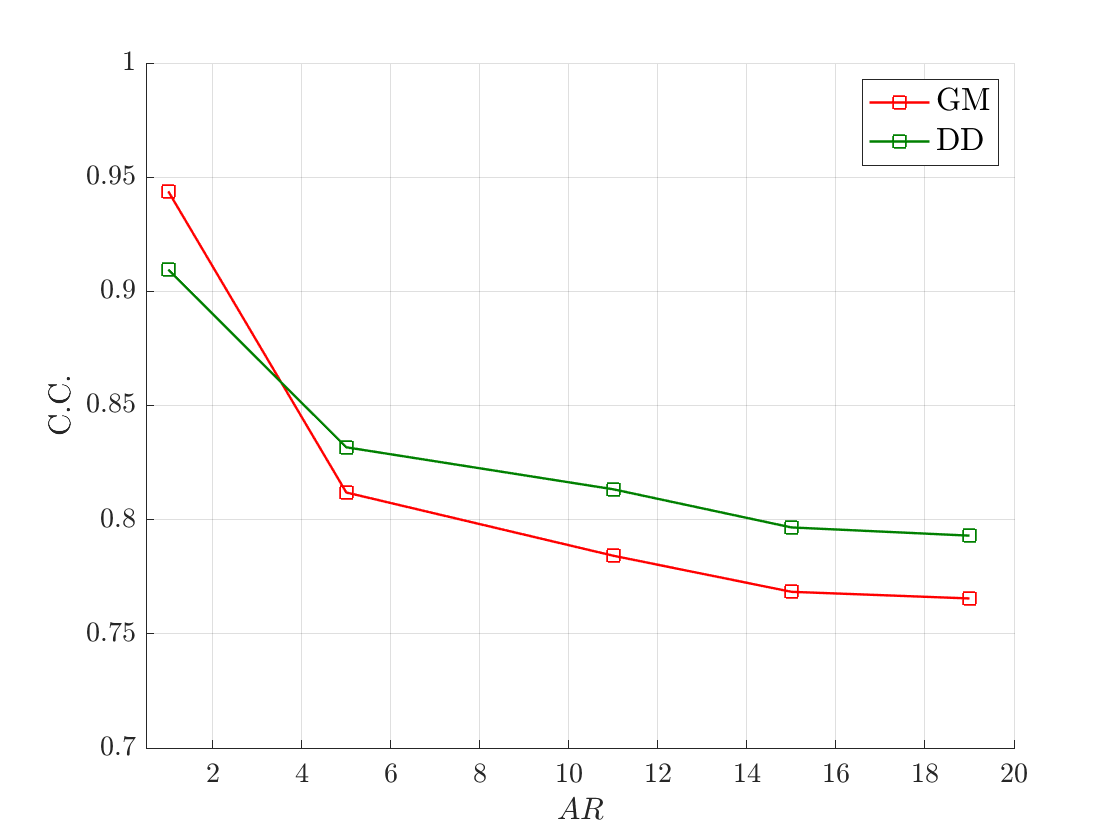}}
    \subfigure[\label{fig:REF_Apriori_Pencil_7dx}]{\includegraphics[width=0.49\textwidth]{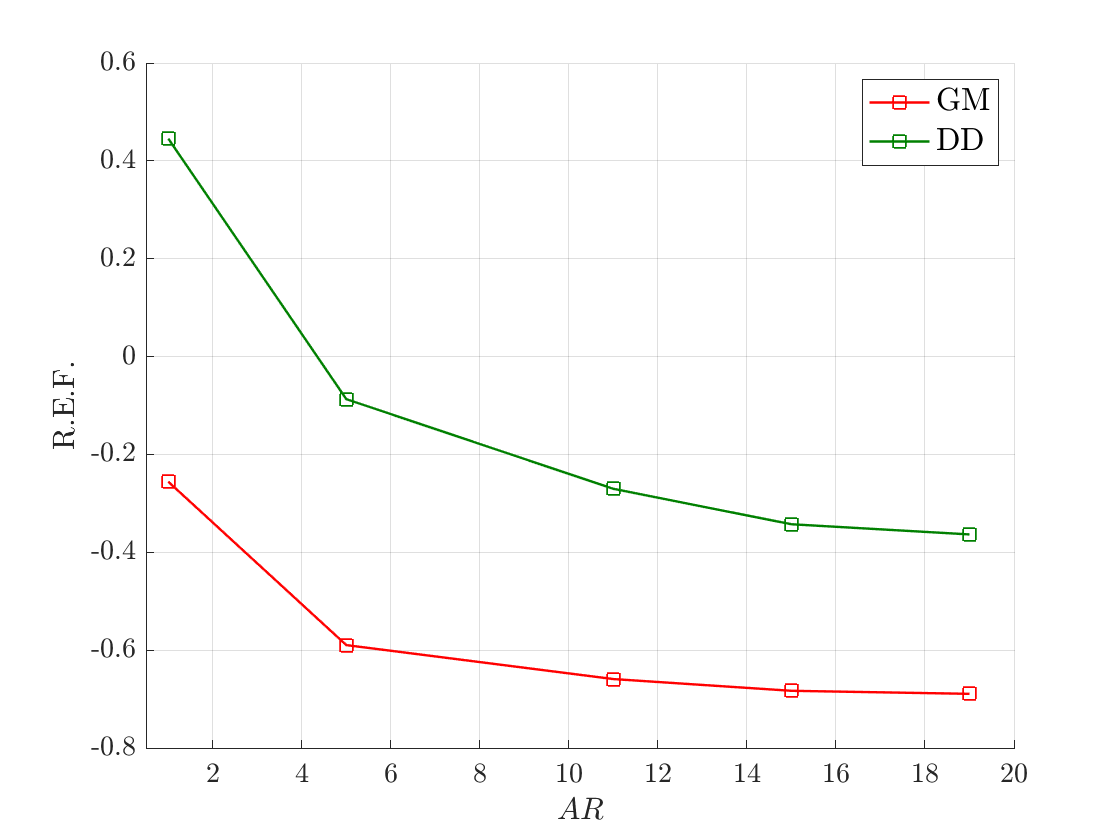}}
    \caption{(a) Correlation Coefficient (C.C.) and (b) Relative error in mean energy flux (R.E.F) for pencil-type anisotropic filters}
    \label{fig:Apriori_Pencil_7dx}
\end{figure}

The results for book-type anisotropic filters for a base filter width of $15.4 \eta$ are shown in \figref{Apriori_Book_7dx}. We observe that for nearly isotropic filters, that is lower values of aspect ratio $AR = \Delta_3/\Delta_1$, stresses predicted by the gradient model have a higher correlation coefficient than those predicted by the data-driven model. However, at higher values of $AR$, the anisotropy of the filter increases and the data-driven model performs better, yielding a higher correlation coefficient of the stresses compared to the gradient model. We observe that R.E.F. predicted by the gradient model rapidly decreases to large negative values with the increase in anisotropy. For the data-driven model, at lower anisotropies, R.E.F is positive. However, with an increase in anisotropy, R.E.F. becomes negative but stays significantly greater than R.E.F. predicted by the gradient model. This behavior indicates that for the case under consideration, the data-driven gives a more accurate prediction of model dissipation than the gradient model for anisotropic filters.

The results for pencil-type anisotropic filters for a base filter width of $15.4 \eta$ are shown in \figref{Apriori_Pencil_7dx}. We observe that the reduction in C.C. with anisotropy for the pencil filter type is more than that observed for book-type anisotropic filters. The data-driven model exhibits a higher correlation coefficient of the SGS stress tensor to the exact DNS SGS tensor than the gradient model for higher anisotropies. Furthermore, the data-driven model also gives better R.E.F. predictions. From the point of view of \textit{a posteriori} simulations, a significant negative R.E.F. relates to energy pile-up at higher wavenumbers and possibility of finite-time numerical instability-induced divergence of simulations. Better R.E.F. predictions for the data-driven points to a better \textit{a posteriori} dissipative performance than the gradient model.

We also considered other base filter widths: $6.6 \eta$ and $28.6 \eta$, for the same aspect ratio of anisotropic filters for the base filter width of $15.4 \eta$. For these filter widths, we observe similar trends in results as data-driven gives a superior correlation coefficient for modeled SGS stresses and better dissipative performance than the gradient model at higher anisotropies. Furthermore, the aspect ratio at which the data-driven model gives a better correlation coefficient than the gradient model seems to decrease at higher base filter widths which is the same behavior as we observed for the isotropic form of the data-driven model in \citep{Prakash2021}.

From this \textit {a priori} test, we observed that the data-driven model works well for anisotropies greater than those in the training set and book-type anisotropic filters that were not a part of the training set.
Therefore, we conclude that the data-driven model trained using limited anisotropy ratios for pencil-type filters appears to generalize well for data outside the training dataset. These results indicate that the proposed model form does not require a large amount of training data to account for the effect of anisotropy in predicted SGS stresses. Instead of learning filter anisotropy, embedding filter anisotropy in the model form allows the model to generalize well to arbitrary anisotropic filters without including them in the training dataset.

\subsection{\textit{A posteriori} results}

\textit{A posteriori} tests were performed using PHASTA which is a stabilized finite element-based computational fluid dynamics (CFD) solver. For the simulations in this article, we use piecewise tri-linear polynomial basis functions for hexahedral grid elements. The generalized-$\alpha$ method is used for temporal discretization \cite{Jansen2000}. The numerical method uses SUPG/PSPG/grad-div stabilization for adjusting to the instabilities arising from pressure-velocity coupling and advective flow. We solve for the advective form of filtered Navier-Stokes equations and use the stabilization matrix formulation mentioned in \cite{Tejada2005}. The code has been validated for several scale-resolving simulations such as LES \cite{Tejada2002} \cite{Tejada2003} \cite{Prakash2021} and DNS \cite{Trofimova2009} \cite{Balin2021}. The dynamic Smagorinsky model employs averaging in homogeneous directions to address common stability issues.

\subsubsection{Forced HIT at $Re_{\lambda} = \infty$}

We first conduct \textit{a posteriori} tests on the flow with the same flow physics as the training dataset and evaluate the ability of the model to generalize to higher Reynolds numbers by considering a high Reynolds number: $Re_{\lambda} = \infty$. The Reynolds number is achieved by setting viscosity to a nominal value of $1\text{x}10^{-12}$. The domain for this case is a cubic box with a side length of $2\pi$. Periodic boundary conditions are applied to each face of the domain. The filter width for the models is the same as the grid size. The flow is initialized with sub-sampled and interpolated instantaneous turbulent flow velocity and pressure distributions obtained from $Re_{\lambda} = 418$ test case from the JHTDB \cite{Li2008}. Forcing is used to inject energy at low wavenumbers for sustaining turbulence. The details of the forcing can be found in \cite{Bazilevs2007} \cite{Prakash2021}. Due to extremely high $Re_{\lambda}$ for this case, DNS data is unavailable. Therefore, we compare the results to the theoretical K41 three-dimensional energy spectra:

\begin{equation}
    E(\kappa) = C \epsilon^{2/3} \kappa^{-5/3},
\end{equation}

\noindent where $C = 1.6$ is the constant determined from theoretical or empirical studies \cite{Pope2000} and $\epsilon$ is turbulence dissipation equal to the power input from forcing at the statistically stationary state. 

\begin{figure}[t!]
    \centering
    \subfigure[No Model\label{fig:Iso_NM_3D}]{\includegraphics[width=0.49\textwidth]{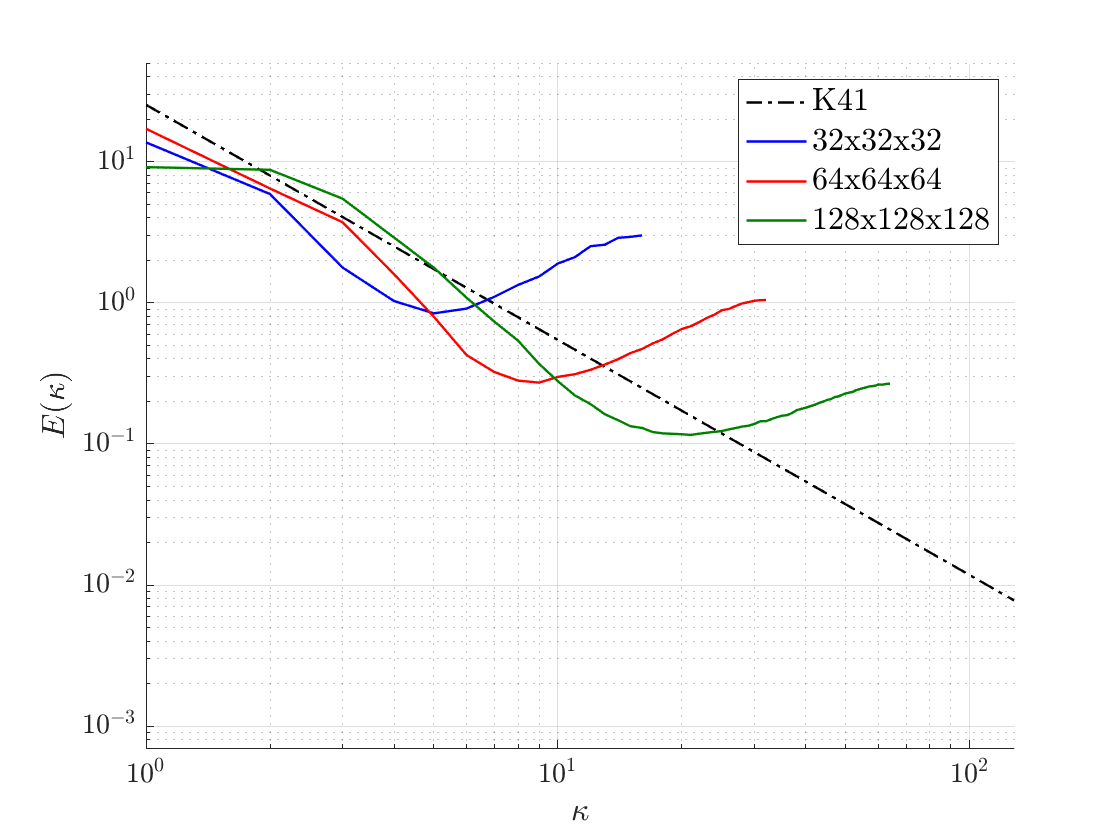}}
    \subfigure[Dynamic Smagorinsky Model\label{fig:Iso_DS_3D}]{\includegraphics[width=0.49\textwidth]{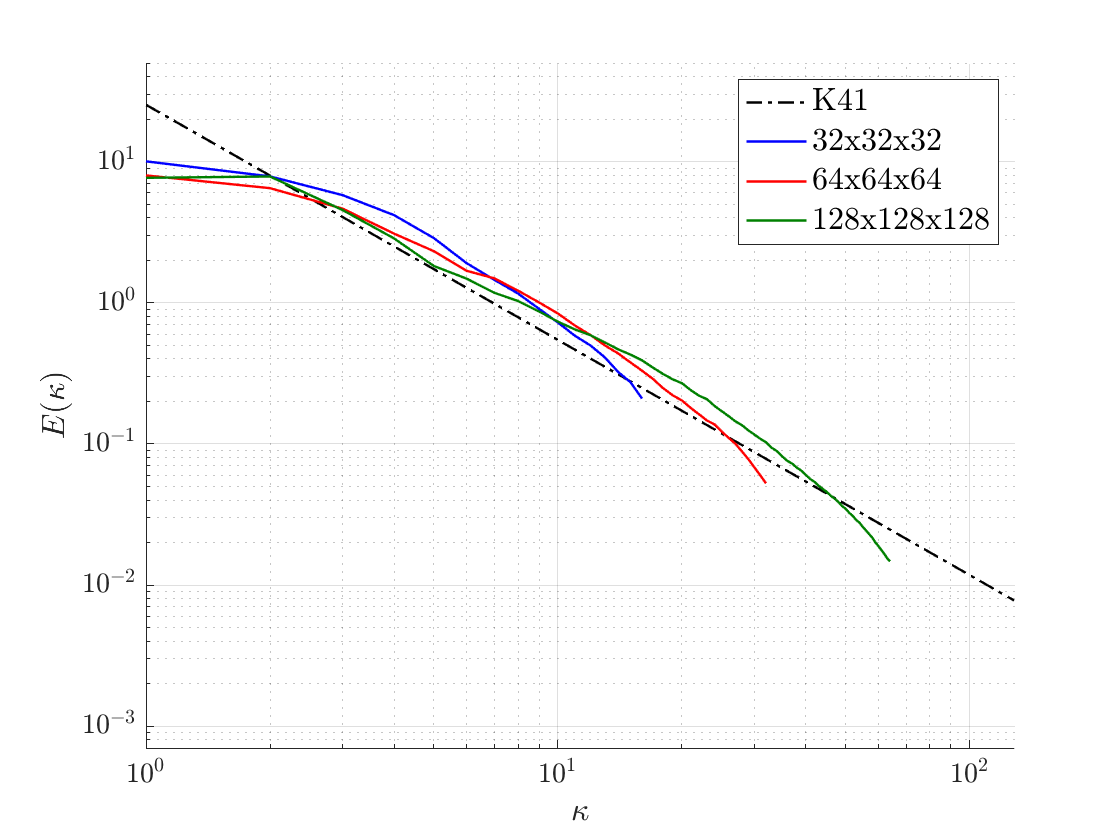}}
    \subfigure[Gradient Model\label{fig:Iso_AGM_NC_3D}]{\includegraphics[width=0.49\textwidth]{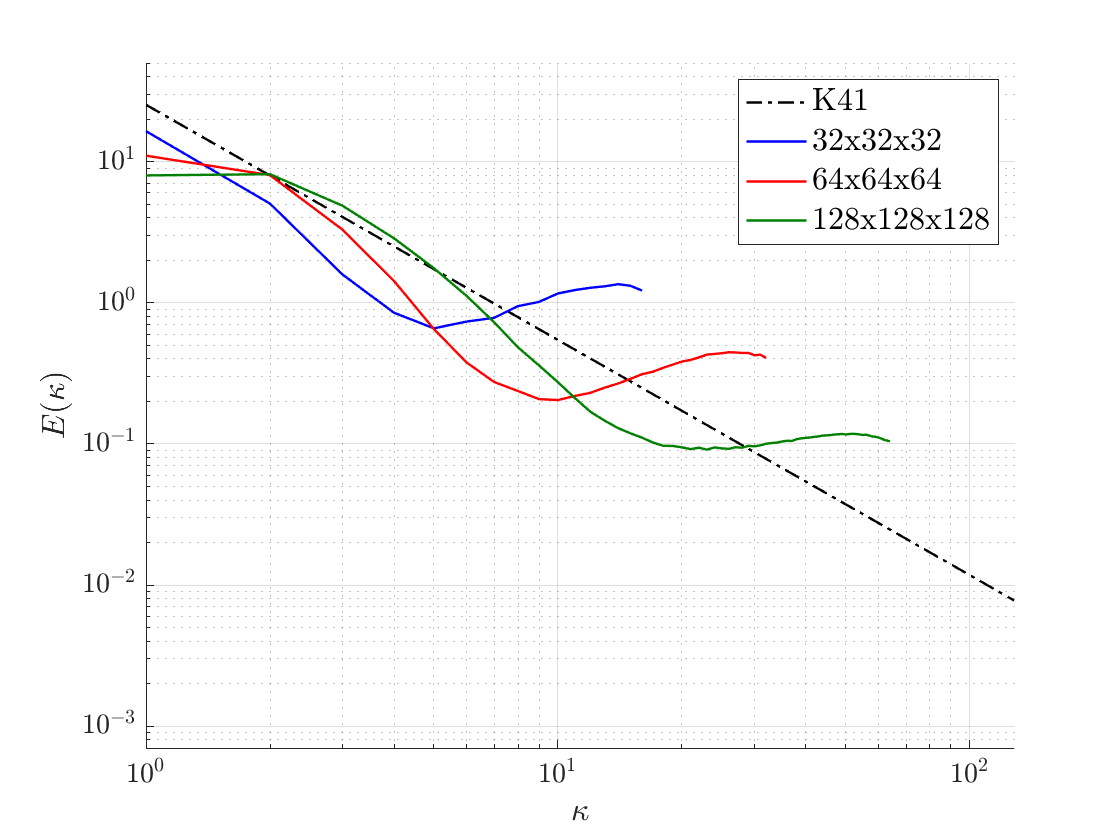}}
    \subfigure[Data-driven Model \label{fig:Iso_ADD_NC_3D}]{\includegraphics[width=0.49\textwidth]{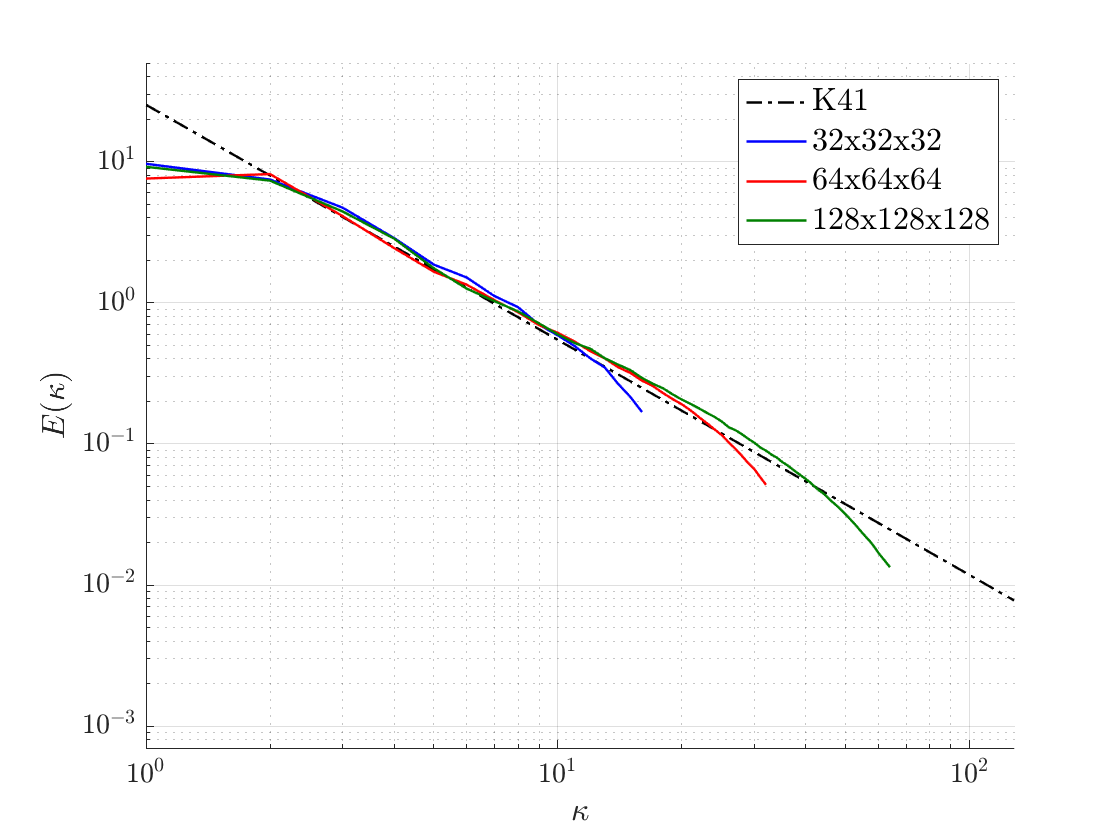}}
    \caption{Energy spectra for isotropic grid resolutions for forced HIT at $Re_{\lambda} = \infty$ }
    \label{fig:ReInf_ES_Iso}
\end{figure}

We first assess the performance of the anisotropic data-driven model in the presence of isotropic grid resolutions. The three-dimensional energy spectra for different SGS models are shown in \figref{ReInf_ES_Iso}. Using no explicit SGS model leads to a pileup of energy at higher resolved wavenumbers for all grid resolutions. For the SUPG/PSPG/grad-div stabilization used in the current simulations, a similar behavior was observed in \citep{Prakash2021}. The results in \citep{Prakash2021} indicated that for a low Reynolds number, such as $Re_{\lambda} = 165$, the stabilization provides sufficient numerical dissipation for the meshes with $64^3$ and $128^3$ grid elements. These results indicate that even though numerical dissipation might be adequate to dissipate the energy at the smallest resolved scales for lower Reynolds numbers, it may not be sufficient for higher Reynolds numbers and using an explicit SGS model is preferable in such cases. The gradient model also generates a significant pileup of energy at higher wavenumbers for all grid resolutions. The dynamic Smagorinsky model exhibits better behavior at higher wavenumbers, although it still exhibits overprediction of energy at the intermediate wavenumbers. The data-driven model yields very accurate energy spectra compared to the theoretical results for all grid resolutions. Note that the data-driven model was trained on HIT data at a lower Reynolds number of $Re_{\lambda} = 418$. As the results are in agreement even for HIT at $Re_{\lambda} = \infty$, the data-driven model appears to generalize well to higher Reynolds numbers. Furthermore, the training dataset for the model included only a single isotropic filter width and that too was a small value in the dissipation range. In this test case, all the filter widths are in the inertial range and the data-driven model seems to perform well which indicates that the anisotropic data-driven still maintains high accuracy for isotropic grids.

\begin{figure}[t!]
    \centering
    \subfigure[No Model\label{fig:Book_NM_3D}]{\includegraphics[width=0.49\textwidth]{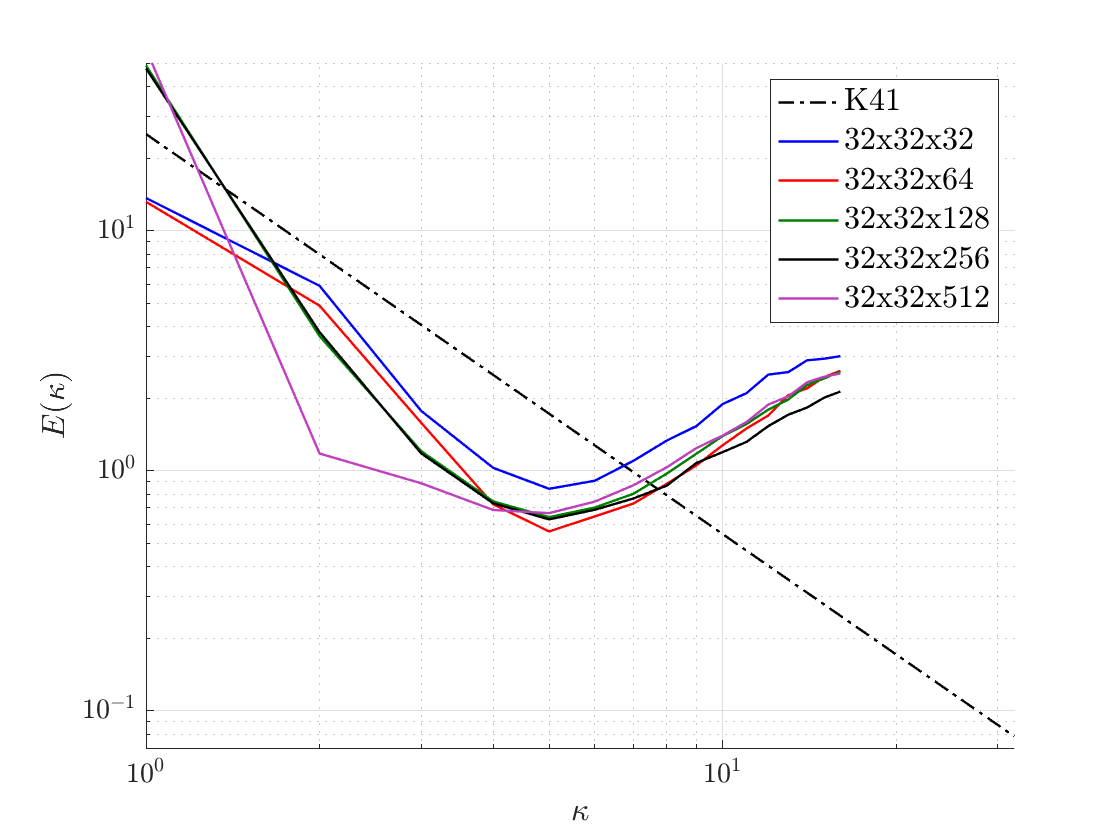}}
    \subfigure[Dynamic Smagorinsky Model\label{fig:Book_DS_3D}]{\includegraphics[width=0.49\textwidth]{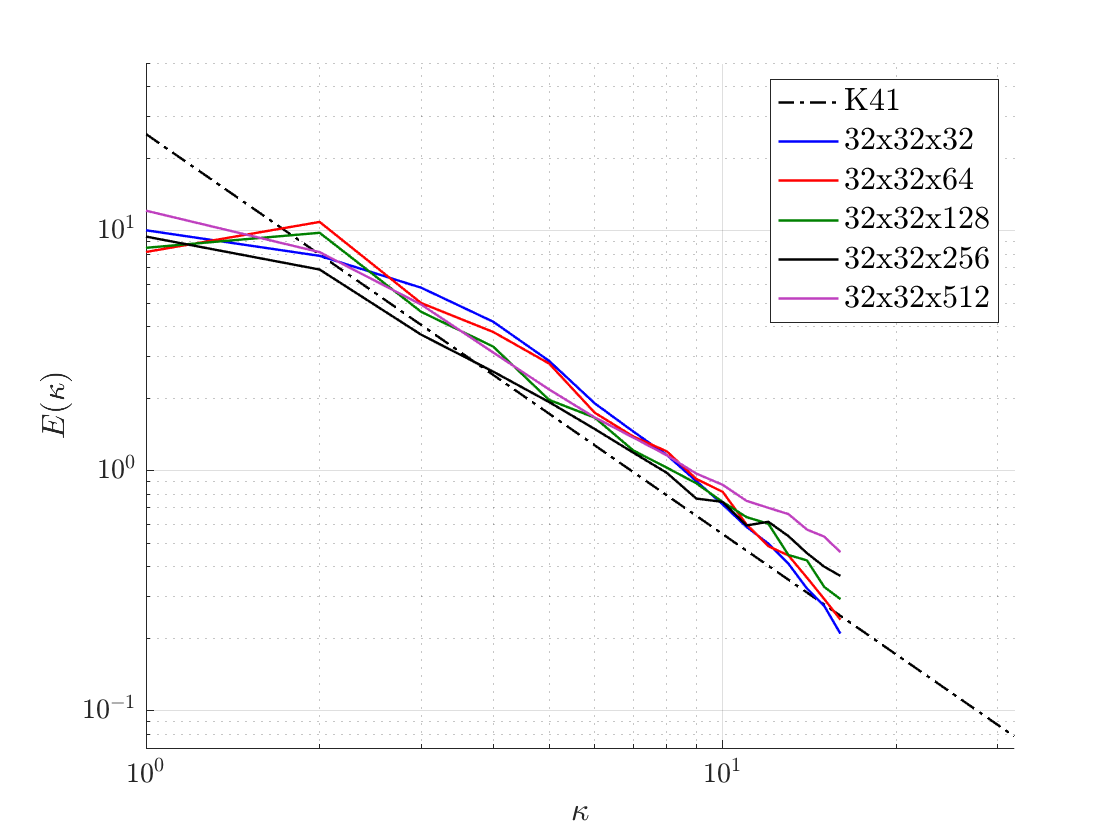}}
    \subfigure[Gradient Model\label{fig:Book_AGM_NC_3D}]{\includegraphics[width=0.49\textwidth]{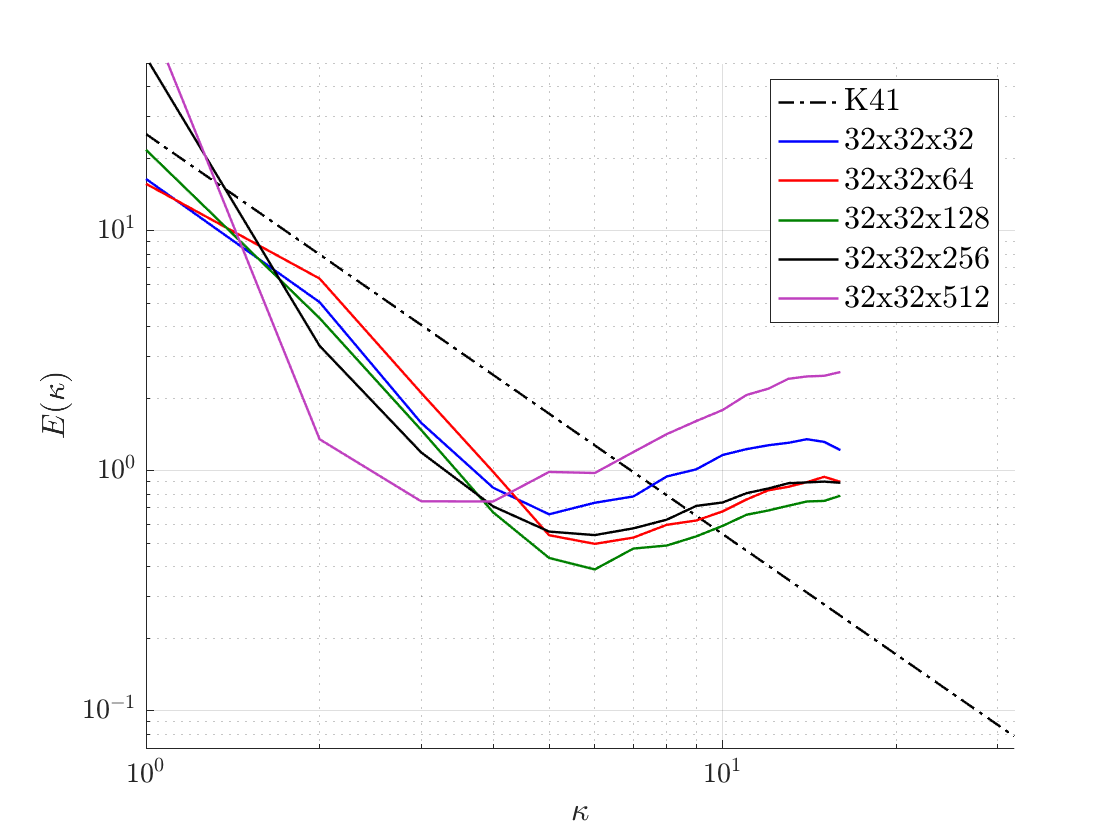}}
    \subfigure[Data-driven Model \label{fig:Book_ADD_NC_3D}]{\includegraphics[width=0.49\textwidth]{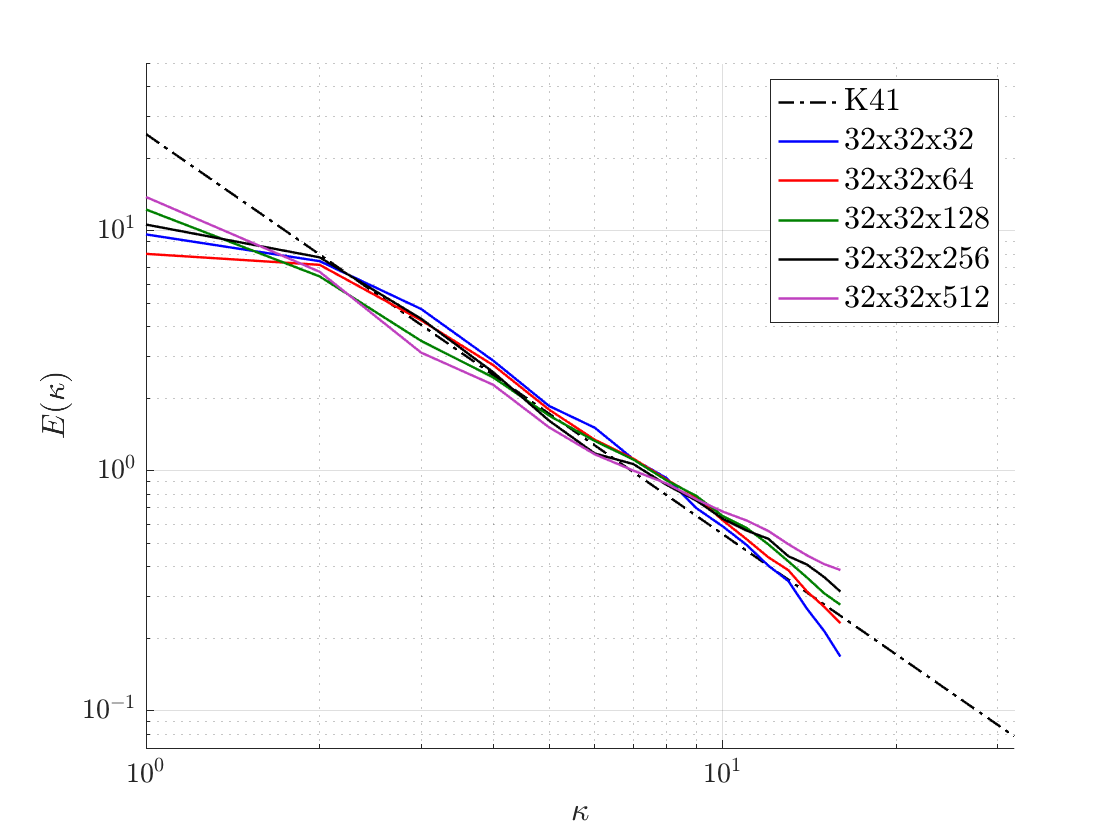}}

    \caption{Energy spectra for book-type grid resolution for forced HIT at $Re_{\lambda} = \infty$ }
    \label{fig:ReInf_ES_Book}
\end{figure}

\begin{figure}[t!]
    \centering
    \subfigure[No Model\label{fig:Pencil_NM_3D}]{\includegraphics[width=0.49\textwidth]{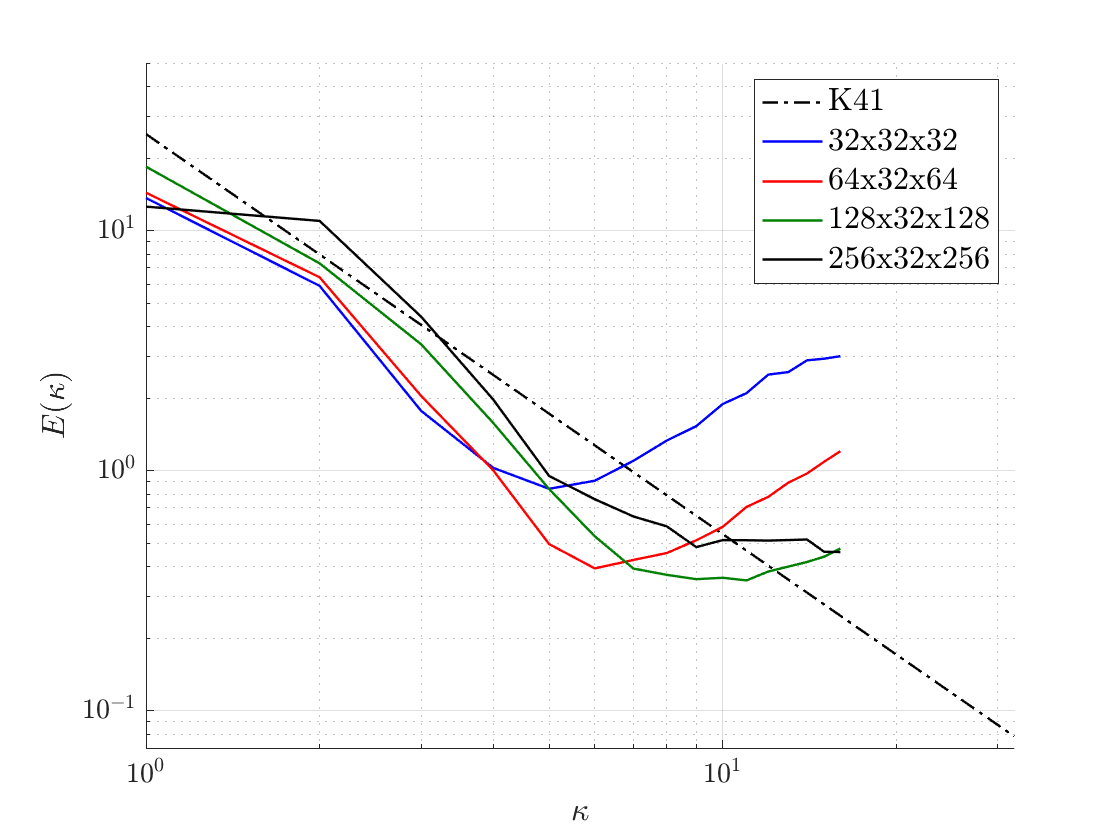}}
    \subfigure[Dynamic Smagorinsky Model\label{fig:Pencil_DS_3D}]{\includegraphics[width=0.49\textwidth]{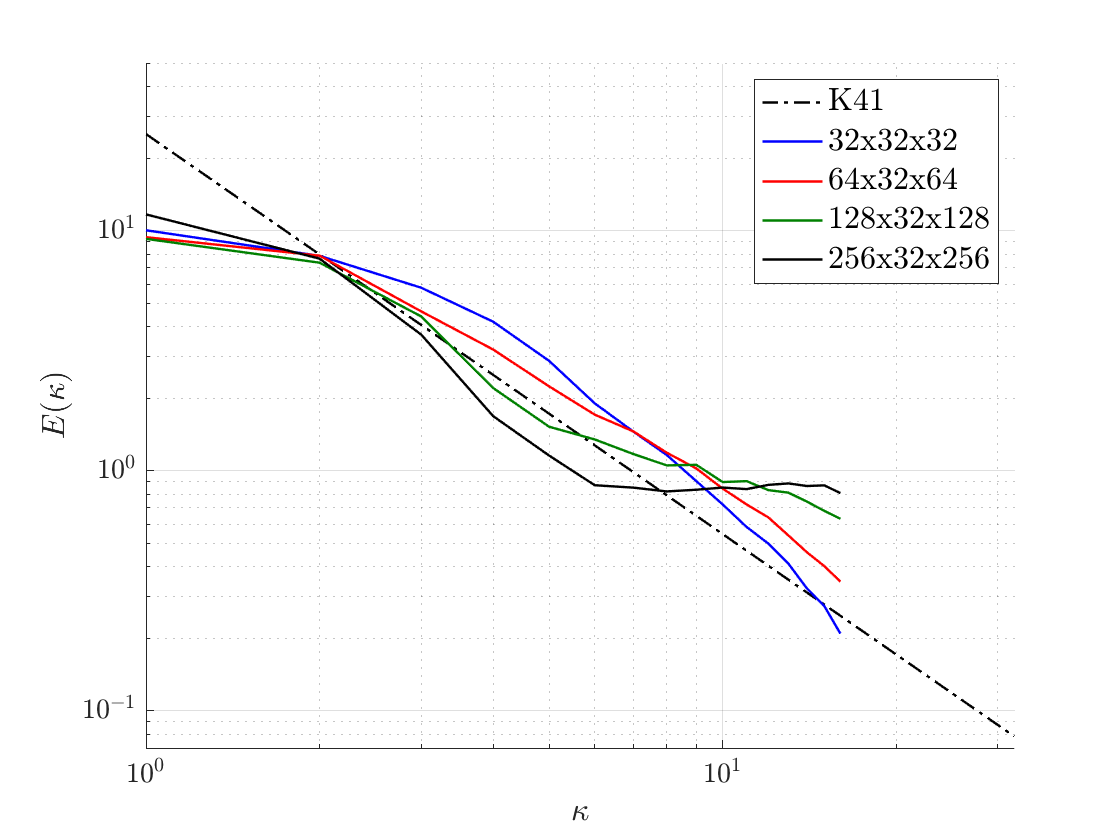}}
    \subfigure[Gradient Model\label{fig:Pencil_AGM_NC_3D}]{\includegraphics[width=0.49\textwidth]{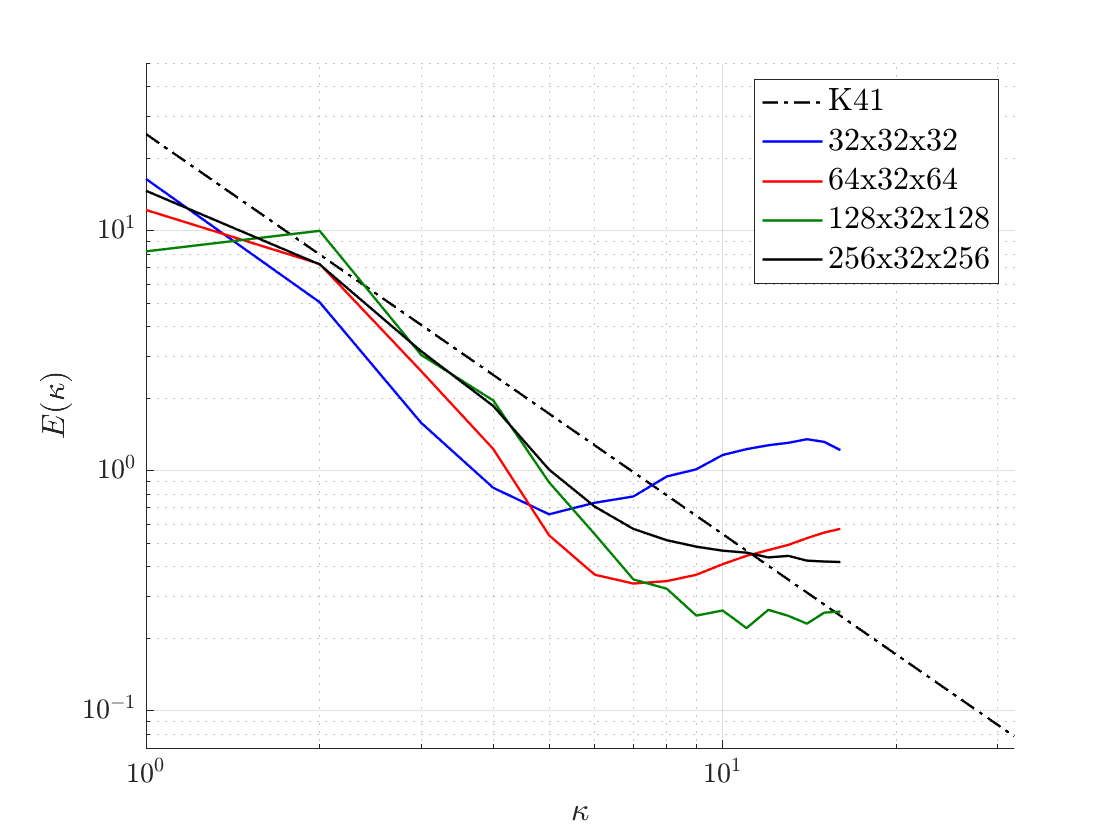}}
    \subfigure[Data-driven Model\label{fig:Pencil_ADD_NC_3D}]{\includegraphics[width=0.49\textwidth]{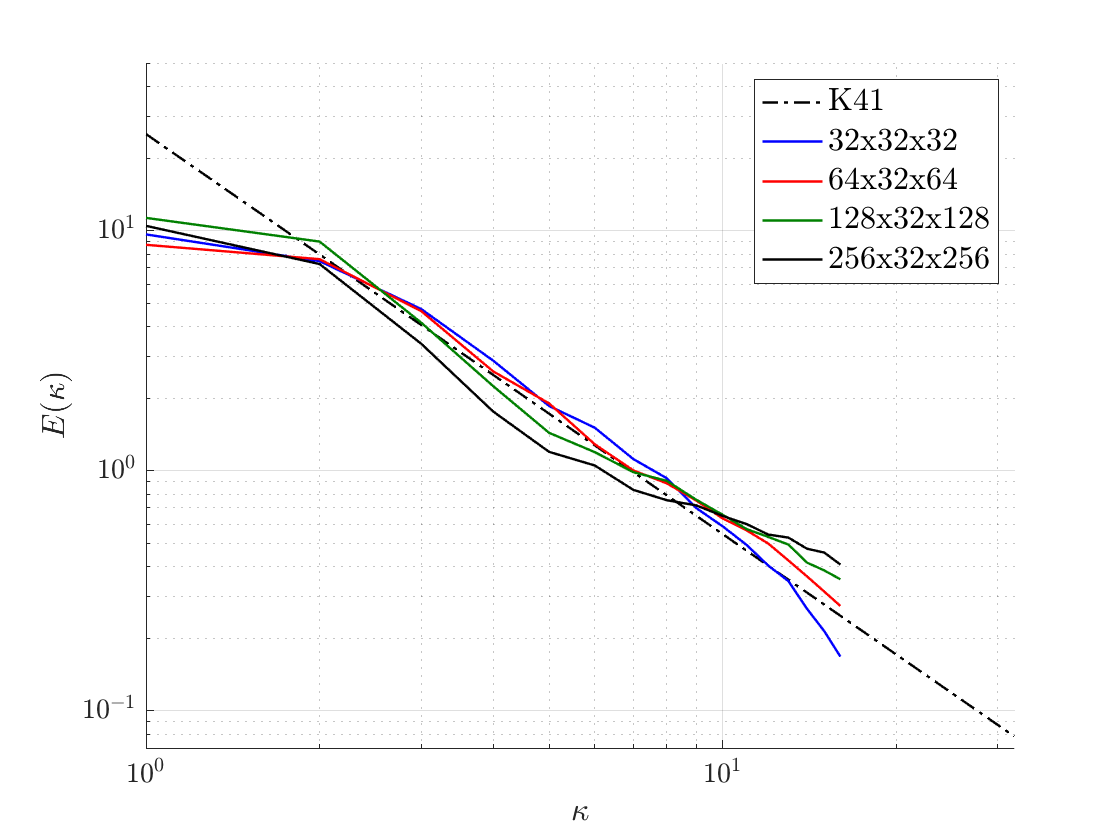}}
    
    \caption{Energy spectra for pencil-type grid resolution for forced HIT at $Re_{\lambda} = \infty$ }
    \label{fig:ReInf_ES_Pencil}
\end{figure}

The three-dimensional energy spectra results for book-type grids are shown in \figref{ReInf_ES_Book}. We observe that using no explicit SGS model or the gradient model results in a large pileup of energy at larger wavenumbers for all isotropic and anisotropic grids. The dynamic Smagorinsky model does not exhibit a pileup of energy, however, it significantly overpredicts the energy at the intermediate wavenumbers for all the resolutions. On the other hand, the data-driven model exhibits better behavior than the other models as the energy spectra are closest to the theoretical results. The results for pencil-type grids are shown in \figref{ReInf_ES_Pencil}. Even for this grid type, the use of no explicit SGS model or the gradient model gives a pileup of energy at the larger wavenumbers. The main distinction compared to book-type grids is that we observe a significant reduction of a pileup of energy as we increase anisotropy. As the grid is refined in multiple directions, more wavenumbers are resolved along those directions, thereby reducing the influence of modeled stresses and dissipation and resulting in a smaller pileup of energy. Alternatively, the behavior is the opposite for the dynamic Smagorinsky and data-driven models. For these models, we observe a slight increase in energy pileup with increasing anisotropy. This behavior highlights that even though the model accounts for anisotropy, very high anisotropy could lead to insufficient model dissipation along the coarse grid direction. Pencil-type grids are not needed for several turbulent flow applications of interest since book-type grids match wall-induced anisotropy, therefore, the slight reduction in accuracy is not of great concern.

The analysis of three-dimensional energy spectra for both pencil and book-type grids shows a consistent pattern in the model performance. The use of no explicit SGS model or the gradient model seems insufficient for all resolutions for the case under consideration. The dynamic Smagorinsky model looks like a better model for simulating this problem. However, we observe that the data-driven model is the best model choice for the model for simulating this problem as it consistently gives better results for all the resolutions and grid-type we considered. Furthermore, the Reynolds number considered for this flow is outside the training dataset. Therefore, the good results obtained using the data-driven model indicate that the model appears to generalize well to Reynolds numbers outside the training dataset. Note that optimal clipping \cite{Prakash2022} can be used to improve the model performance of the gradient model. Using optimal clipping for both gradient and data-driven models results in better predictions, but the same conclusions follow as the data-driven model is superior to the gradient model. In this article, we have not included this discussion for brevity and we only compare results for the original model without the added regularization offered by optimal clipping. 

\subsubsection{Turbulent channel flow at $Re_{\tau} = 395$ and $590$}

We perform wall-resolved LES of turbulent channel flow to demonstrate the performance of the anisotropic data-driven SGS model for wall-bounded flows. As anisotropic grids are mostly used for simulating turbulent flow through a channel, this flow is well suited to demonstrate the applicability of the anisotropic data-driven SGS model. We consider two Reynolds numbers for the turbulent channel flow: $Re_{\tau} = 395$ and $Re_{\tau} = 590$. A domain of $2 \pi \delta \; \times 2 \delta \times \pi \delta$ is used, where $\delta$ ($ = 1 $) is the channel half-height. The flow is periodic in streamwise and spanwise directions. A no-slip wall boundary condition is used at $y=0$ and $y=2 \delta$. The details on grid resolutions for the two Reynolds number cases are mentioned in \tabref{Channel_mesh}, where $\Delta x^+$, $\Delta z^+$, $\Delta y_1^+$ and $\Delta y_c^+$ are the streamwise grid spacing, spanwise grid spacing, wall-normal grid spacing for the first off-wall element and wall-normal grid spacing at the channel centerline, all non-dimensionalized with inner-region units ($\Delta^+ = \Delta u_{\tau}/ \nu$). A constant mass flux forcing, based on bulk Reynolds number ($Re_b$) of  $6800$ and $10975$ for $Re_{\tau}$ of $395$ and $590$ respectively, is used to sustain the flow. The flow is initialized using a log-law velocity profile with added random Gaussian perturbations. After the initial transient period, streamwise-averaged and spanwise-averaged flow statistics are extracted. These statistics are further time-averaged over at least $45 T_f$, where $T_f$ is a single flow-through time. Velocity profiles extracted from the simulations are compared to DNS results presented in \cite{Moser1999}. Similarly, the deviatoric part of the Reynolds stress tensor,

\begin{equation}
    a_{ij} = \langle u'_i u'_j \rangle - \frac{1}{3} \langle u'_i u'_i \rangle,
\end{equation}

\noindent is also extracted and compared to the DNS counterpart. 


\begin{table}[t!]
    \centering
    \begin{tabular}{ccccccc}
        \hline
        \hline
         $Re_{\tau}$ & \textbf{Mesh Resolution} & \textbf{Number of Elements} & \textbf{$\Delta x^+$} & \textbf{$\Delta y_1^+$} & \textbf{$\Delta y_c^+$} & \textbf{$\Delta z^+$} \\
         \hline
         395 & Coarse & 32 $\times$ 85 $\times$ 32 & 77.5 & 1 & 34 & 39 \\
         395 & Fine & 64 $\times$ 119 $\times$ 64 & 39 & 1 & 21 & 20 \\ 
         590 & Coarse & 48 $\times$ 111 $\times$ 48 & 78 & 1 & 45 & 39 \\ 
         590 & Medium & 64 $\times$ 133 $\times$ 64 & 58 & 1 & 36 & 29 \\
         \hline
    \end{tabular}
    \caption{Mesh parameters for the channel flow case}
    \label{tab:Channel_mesh}
\end{table}

\begin{table}[b!]
    \centering
    \begin{tabular}{ccccccc}
        \hline
        \hline
         \textbf{Mesh Resolution} & \textbf{DNS} & \textbf{DS} & \textbf{GM} & \textbf{DD}  \\
         \hline
         Coarse & 0.0066 & 0.0051 & 0.0071 & 0.0062  \\   
         Fine & 0.0066 & 0.0056 & 0.0070 & 0.0067\\  
         \hline
    \end{tabular}
    \caption{Skin-friction coefficient prediction for turbulent channel flow at $Re_{\tau} = 395$}
    \label{tab:Cf_ReT395}
\end{table}

\begin{table}[b!]
    \centering
    \begin{tabular}{ccccccc}
        \hline
        \hline
         \textbf{Mesh Resolution} & \textbf{DNS} & \textbf{DS} & \textbf{GM} & \textbf{DD} \\
         \hline
         Coarse & 0.0058 & 0.0047 & 0.0063 &  0.0055  \\   
         Medium & 0.0058 & 0.0049 & 0.0064 &  0.0058 \\ 
         \hline
    \end{tabular}
    \caption{Skin-friction coefficient prediction for turbulent channel flow at $Re_{\tau} = 590$}
    \label{tab:Cf_ReT590}
\end{table}

We compare the skin-friction coefficient, $C_f = \tau_w/ (\frac{1}{2} \rho \bar{u}_b^2 )$, for turbulent channel flow for $Re_{\tau} = 395$ and $Re_{\tau} = 590$ in \tabref{Cf_ReT395} and \tabref{Cf_ReT590} respectively. For the flow at $Re_{\tau} = 395$, we observe that the dynamic Smagorinsky model undepredicts $C_f$ for both grid resolutions. The gradient model results in an overprediction of $C_f$, whereas the data-driven model gives the closest prediction to DNS for both grid resolutions. We observe similar behavior for $Re_{\tau} = 590$ and the data-driven model gives the closest prediction to the DNS results.

\begin{figure}[t!]
    \centering
    \subfigure[\label{fig:395_up_coarse}]{\includegraphics[width=0.49\textwidth]{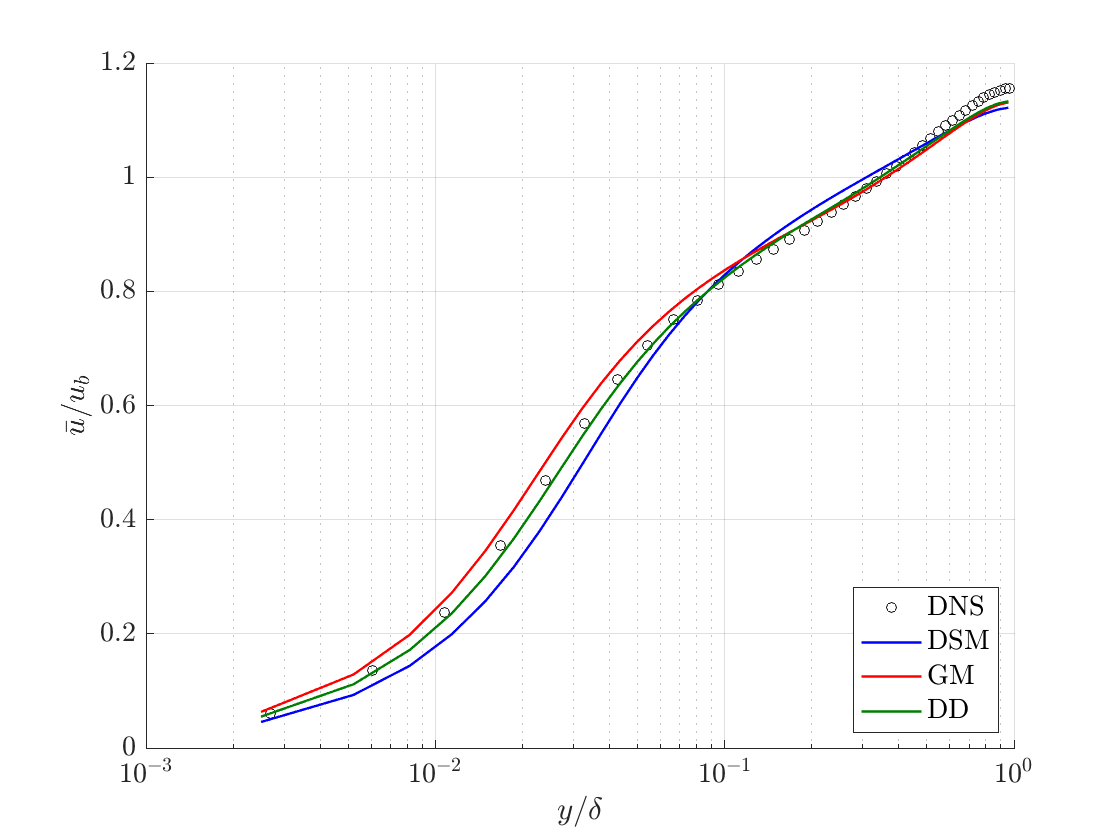}}
    \subfigure[\label{fig:395_up_fine}]{\includegraphics[width=0.49\textwidth]{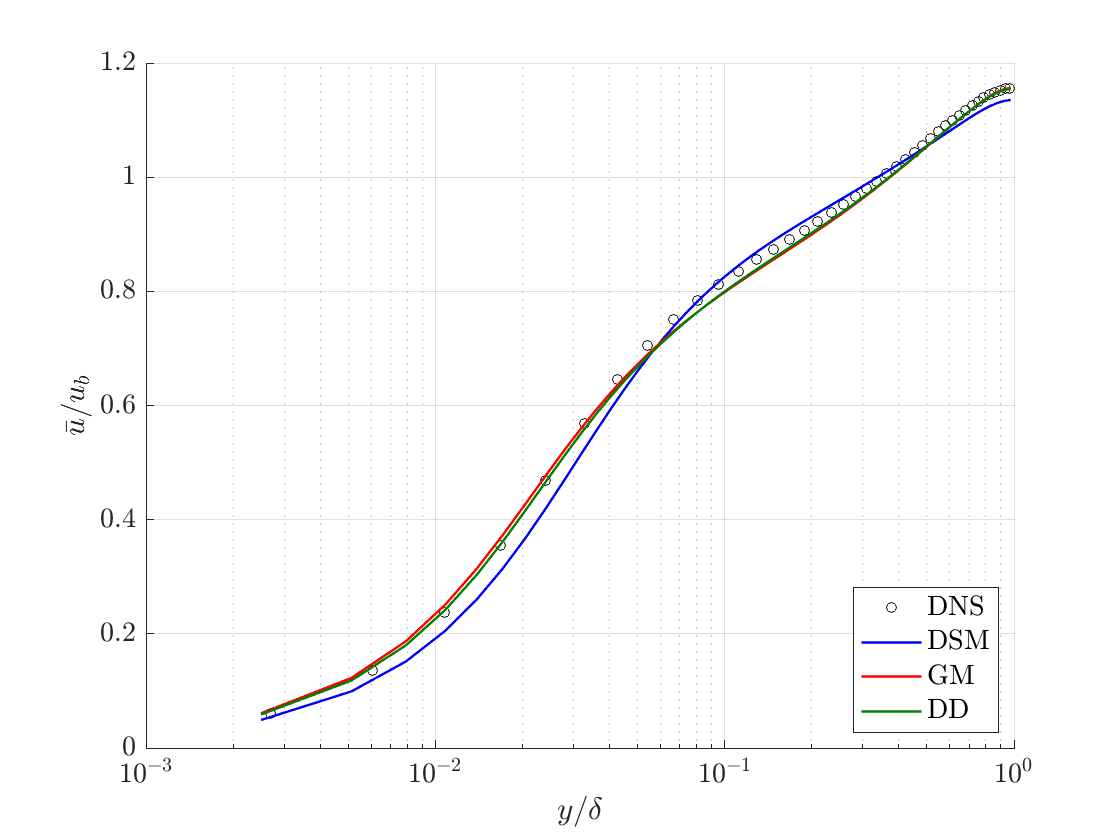}}    
    \caption{Velocity profiles for the a) coarse grid resolution and b) fine grid resolution for turbulent channel flow at $Re_{\tau} = 395$}
    \label{fig:Channel_395_up}
\end{figure}

The mean velocity profiles for $Re_{\tau} = 395$ for both coarse and fine grid resolution are shown in \figref{Channel_395_up}. Instead of using friction velocity ($u_{\tau}$), we use bulk velocity ($u_{b}$) to scale the velocity profile to avoid the effect of chosen scaling on the scaled profiles. The bulk velocity remains the same due to mass forcing and is better suited for scaling velocity and stress profiles. For the coarse grid resolution, we observe that the dynamic Smagorinsky model leads to a significant underprediction of results close to the wall. The gradient model and the data-driven model give the closest prediction of the mean velocity profile to the DNS with the latter model giving slightly better results. For the fine grid resolution, predictions by the gradient model and the data-driven model are closer to the DNS, whereas the dynamic Smagorinsky model underpredicts the mean velocity profile close to the wall.

\begin{figure}[t!]
    \centering
    \subfigure[\label{fig:395_uu_coarse}]{\includegraphics[width=0.49\textwidth]{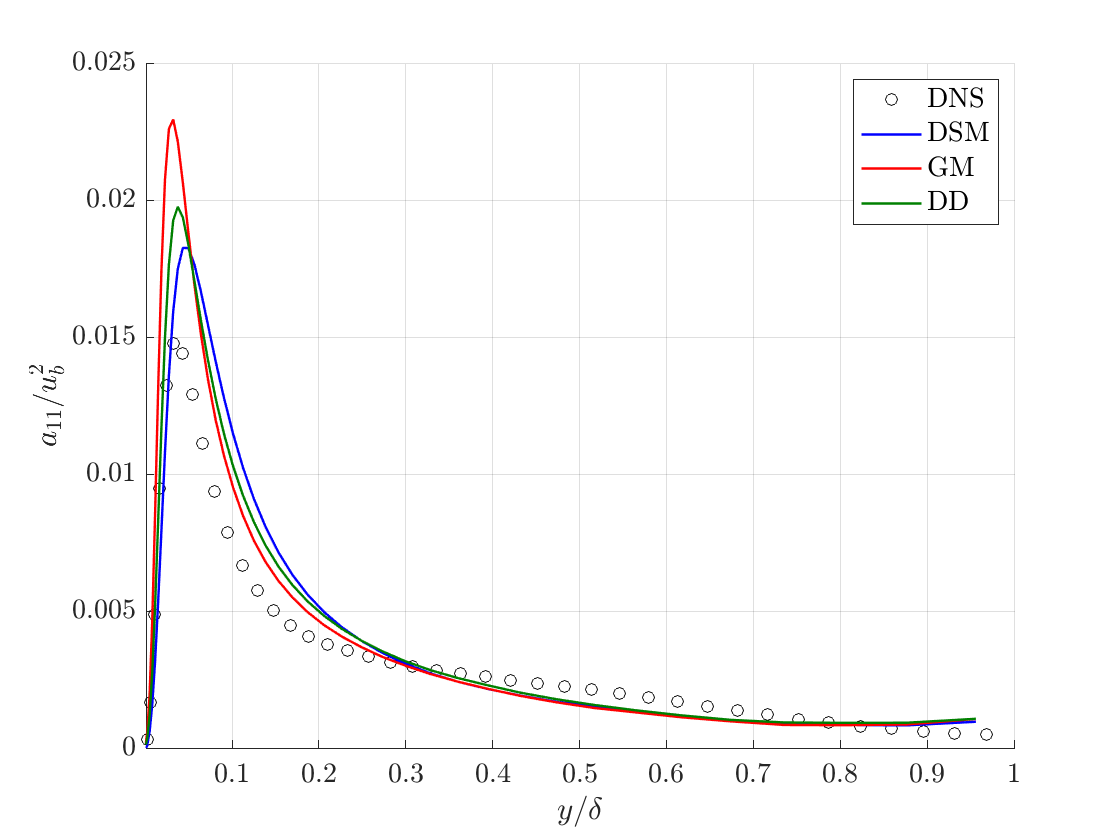}}
    \subfigure[\label{fig:395_vv_coarse}]{\includegraphics[width=0.49\textwidth]{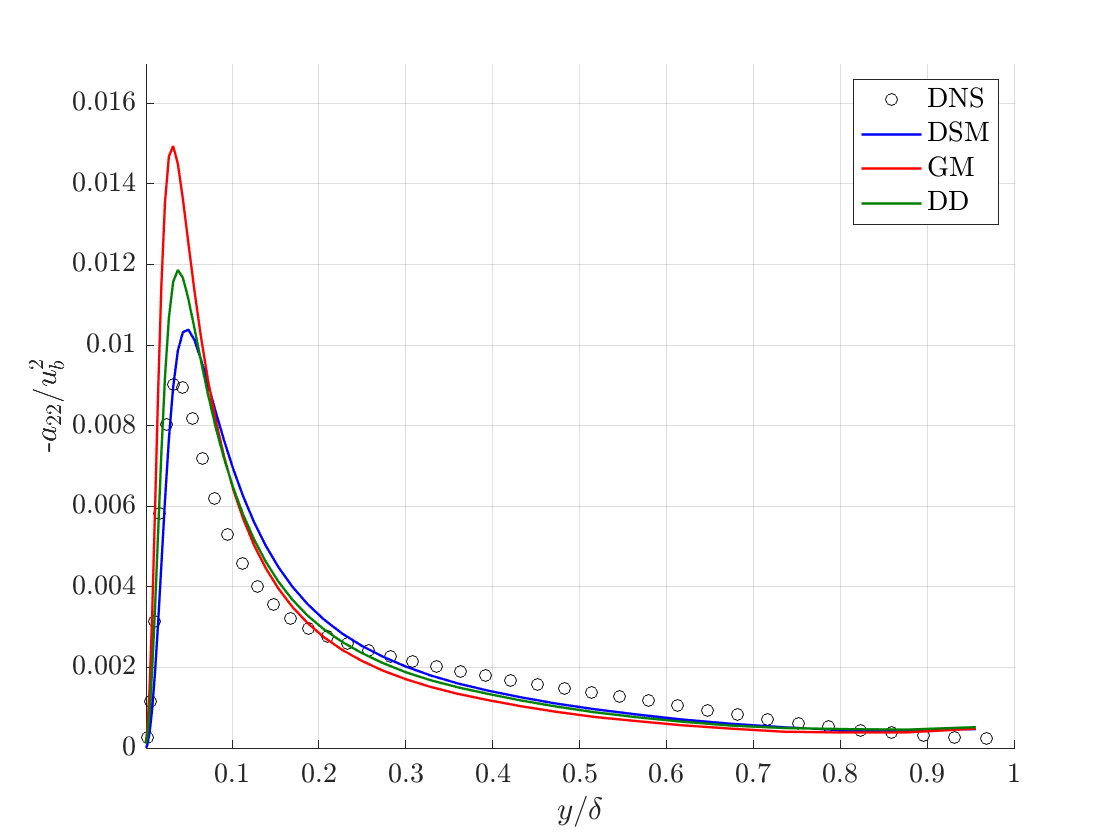}}
    \subfigure[\label{fig:395_ww_coarse}]{\includegraphics[width=0.49\textwidth]{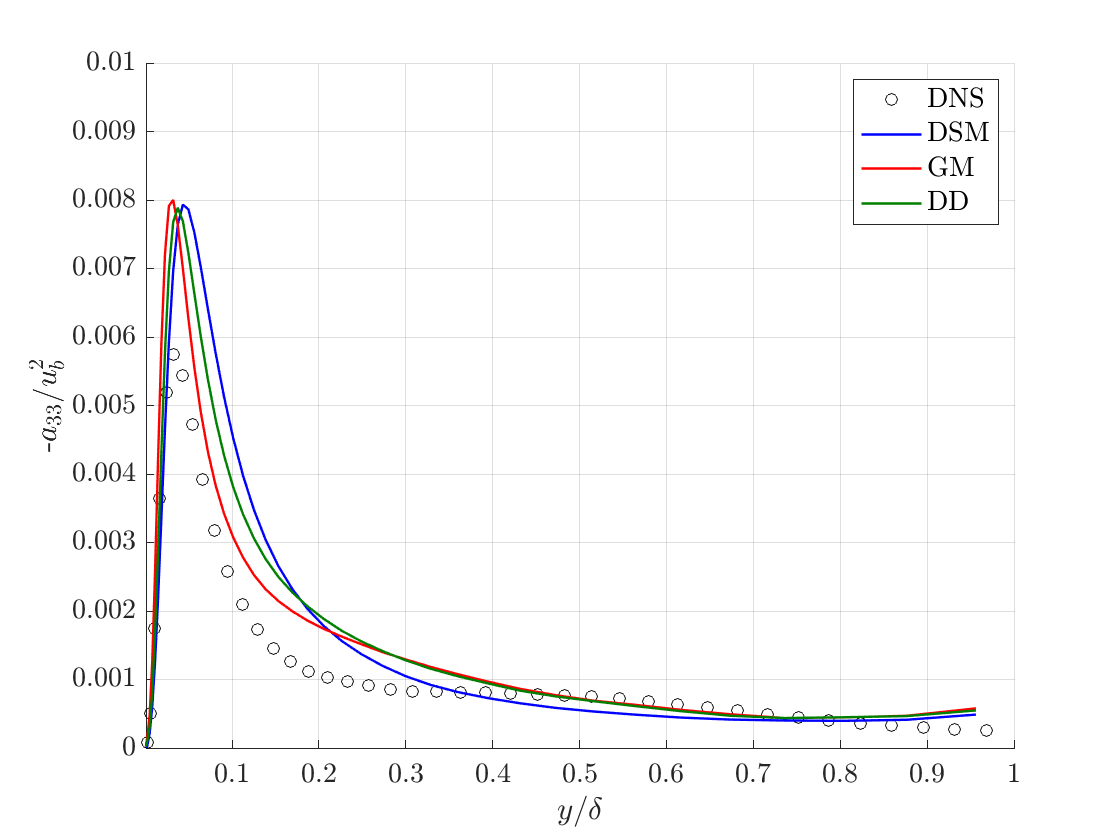}}
    \subfigure[\label{fig:395_uv_coarse}]{\includegraphics[width=0.49\textwidth]{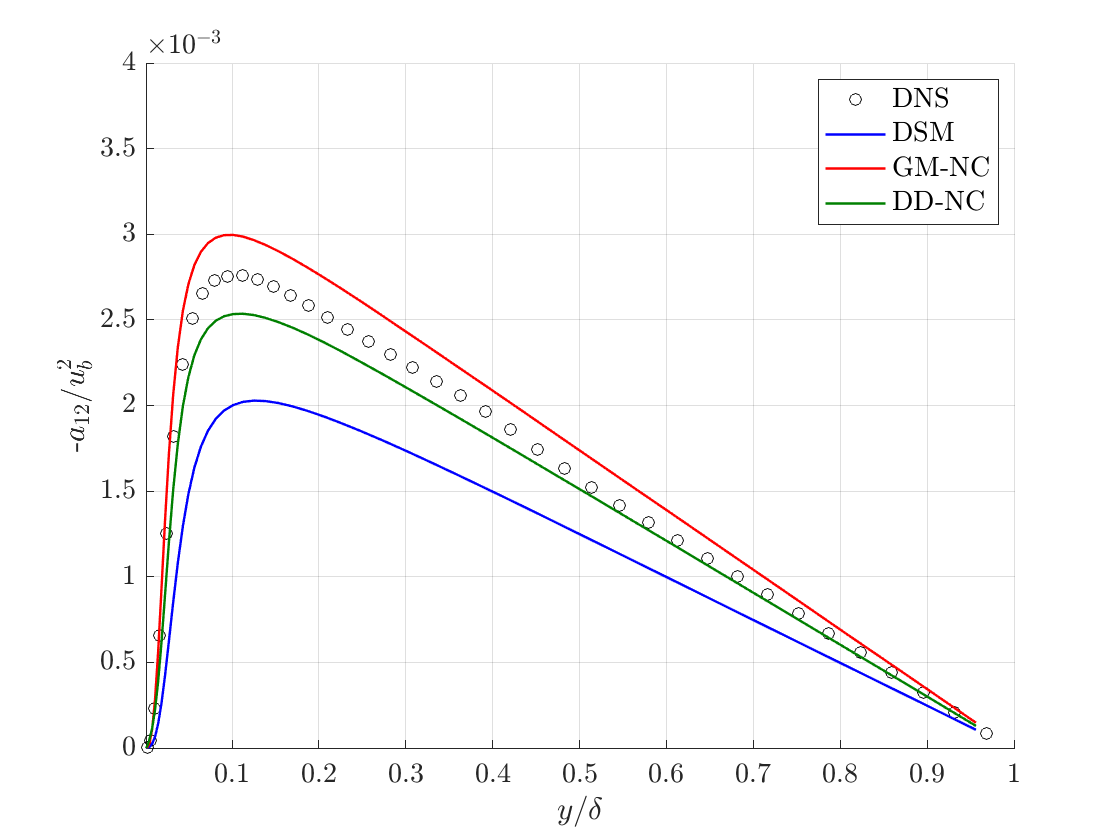}}    
    \caption{ (a) $a_{11}$, (b) $a_{22}$, (c) $a_{33}$ and (d) $a_{12}$ for the coarse grid resolution for turbulent channel flow at $Re_{\tau} = 395$}
    \label{fig:Channel_395_coarse}
\end{figure}

\begin{figure}[t!]
    \centering
    \subfigure[\label{fig:395_uu_fine}]{\includegraphics[width=0.49\textwidth]{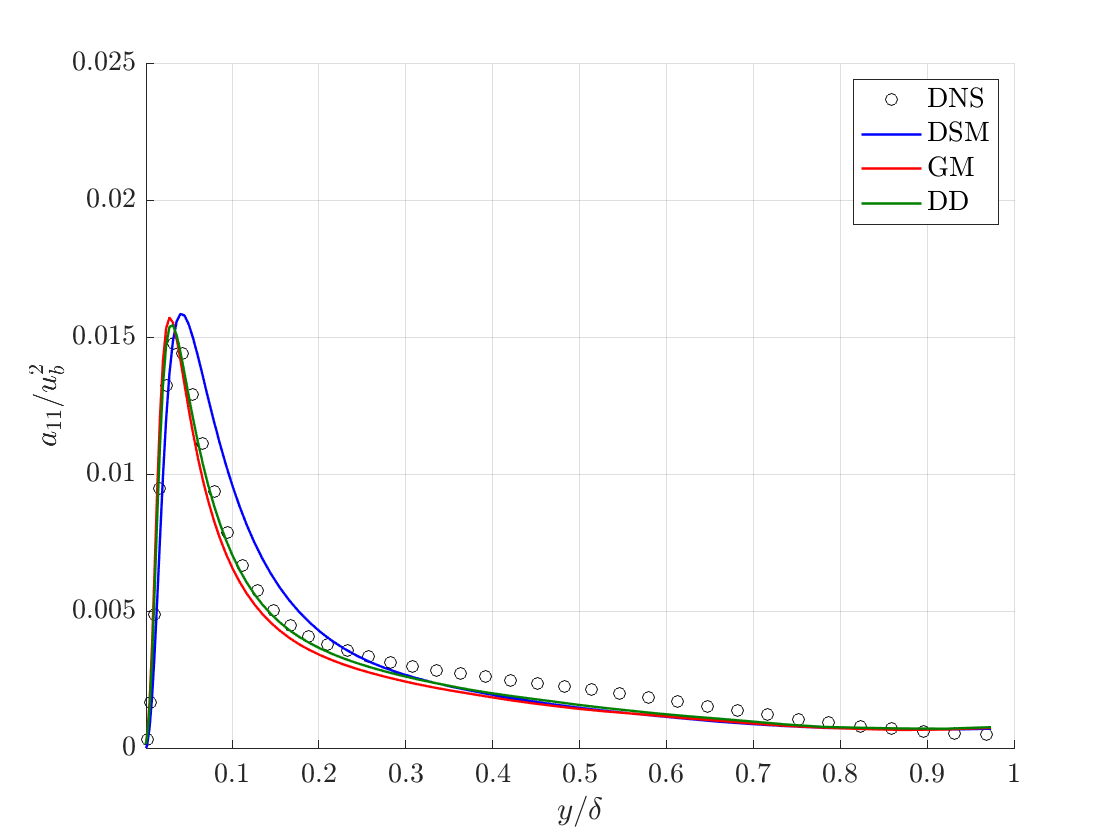}}
    \subfigure[\label{fig:395_vv_fine}]{\includegraphics[width=0.49\textwidth]{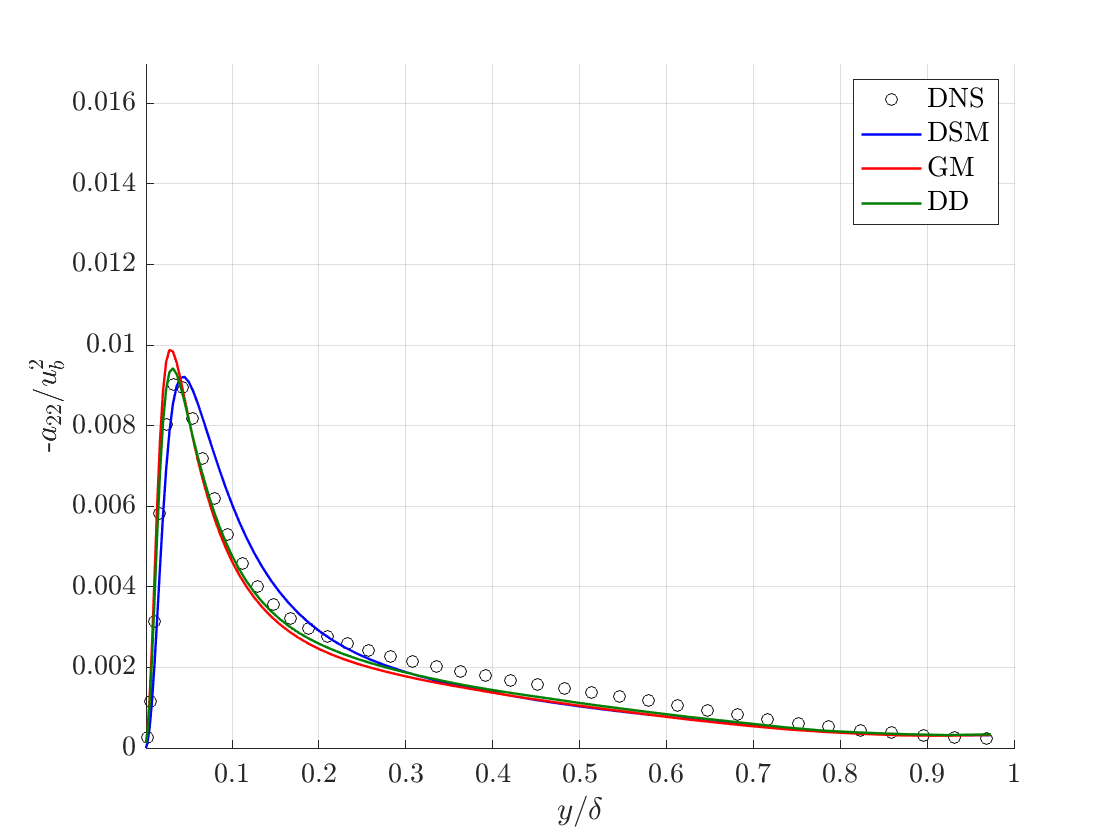}}
    \subfigure[\label{fig:395_ww_fine}]{\includegraphics[width=0.49\textwidth]{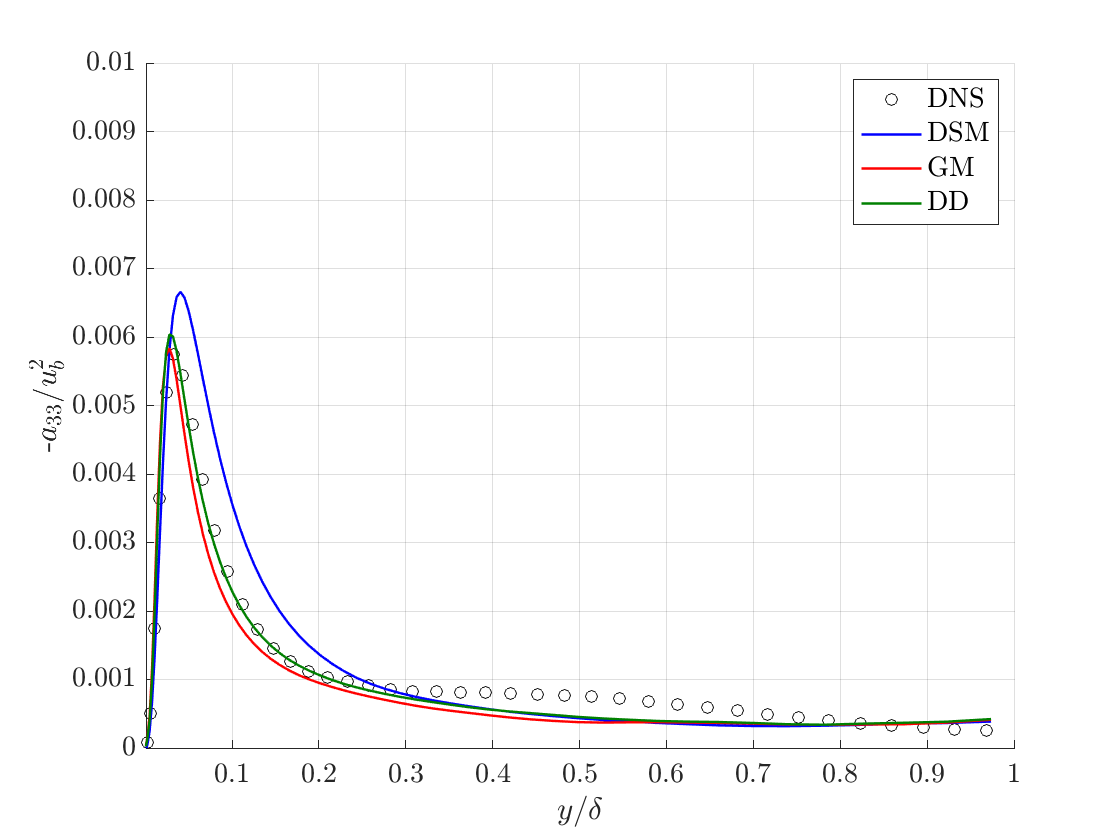}}
    \subfigure[\label{fig:395_uv_fine}]{\includegraphics[width=0.49\textwidth]{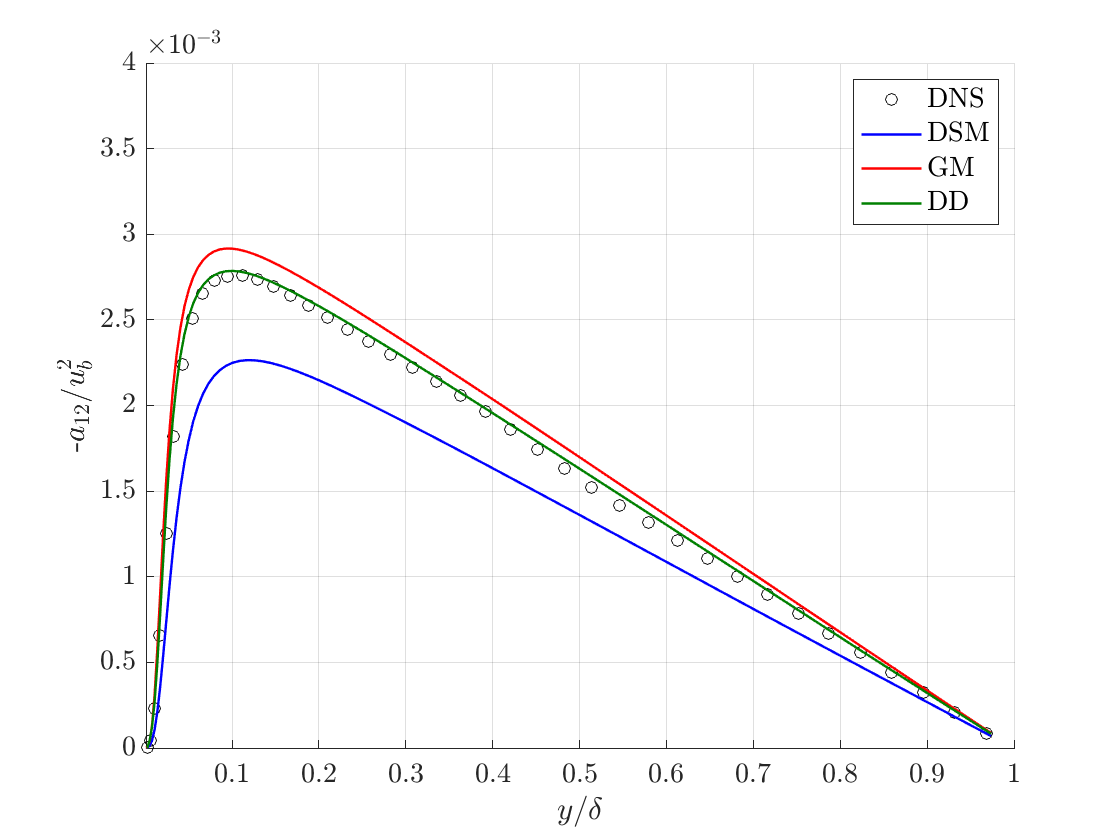}}    
    \caption{ (a) $a_{11}$, (b) $a_{22}$, (c) $a_{33}$ and (d) $a_{12}$ for the fine grid resolution for turbulent channel flow at $Re_{\tau} = 395$}
    \label{fig:Channel_395_fine}
\end{figure}

The components of the deviatoric part of the Reynolds stresstensor for the coarse and fine grid resolutions for turbulent channel flow at $Re_{\tau} = 590$ are shown in \figref{Channel_395_coarse} and \figref{Channel_395_fine} respectively. We observe that all explicit SGS models overpredict the peak value of normal stresses for the coarse grid resolution. The dynamic Smagorinsky model significantly underpredicts the Reynolds shear stresses. On the other hand, data-driven and gradient models give closer results to the DNS with the former underpredicting the results slightly and the latter slightly overpredicting them. For the fine grid resolution, the dynamic Smagorinsky model slightly overpredicts the peak normal stress in the streamwise and spanwise directions. The gradient model overpredicts the peak wall normal stresses. The data-driven model gives the closest prediction of peak normal stresses to the DNS. The gradient model over-predicts the Reynolds shear stress, whereas the data-driven model the data-driven model provides the best Reynolds shear stress prediction with the predictions almost overlapping DNS results.

\begin{figure}[t!]
    \centering
    \subfigure[\label{fig:590_up_coarse}]{\includegraphics[width=0.49\textwidth]{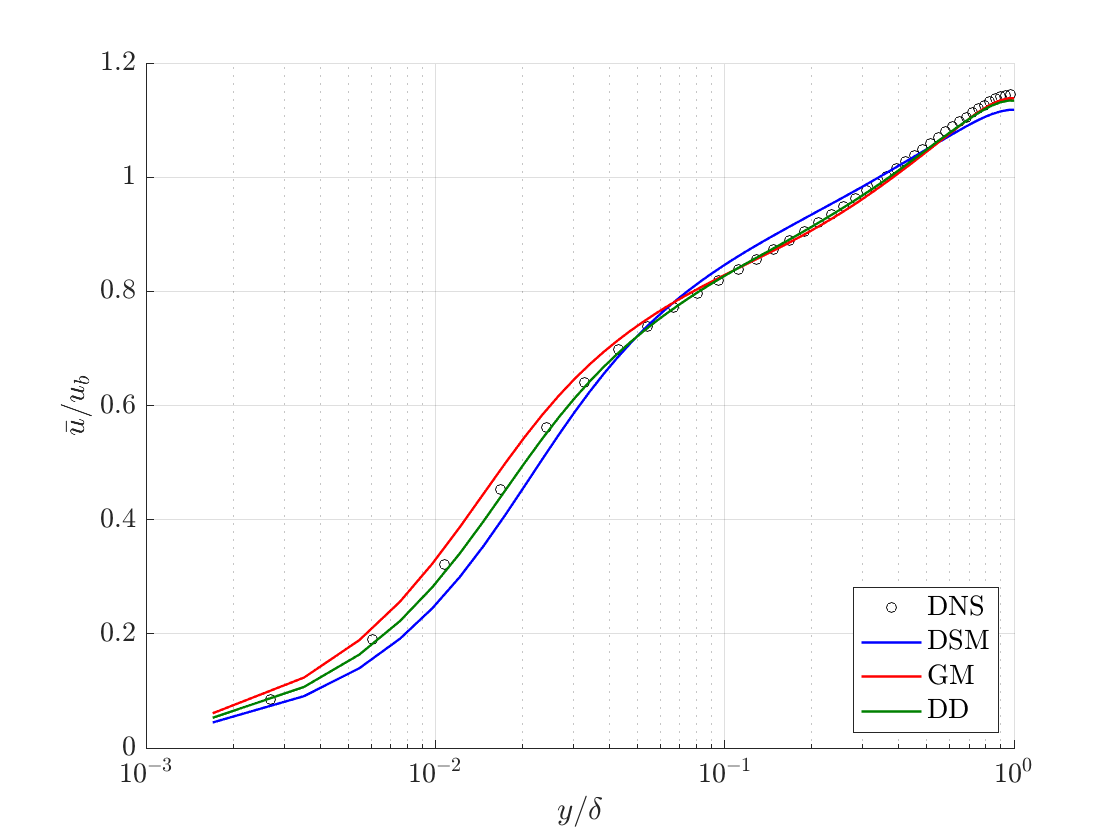}}
    \subfigure[\label{fig:590_up_fine}]{\includegraphics[width=0.49\textwidth]{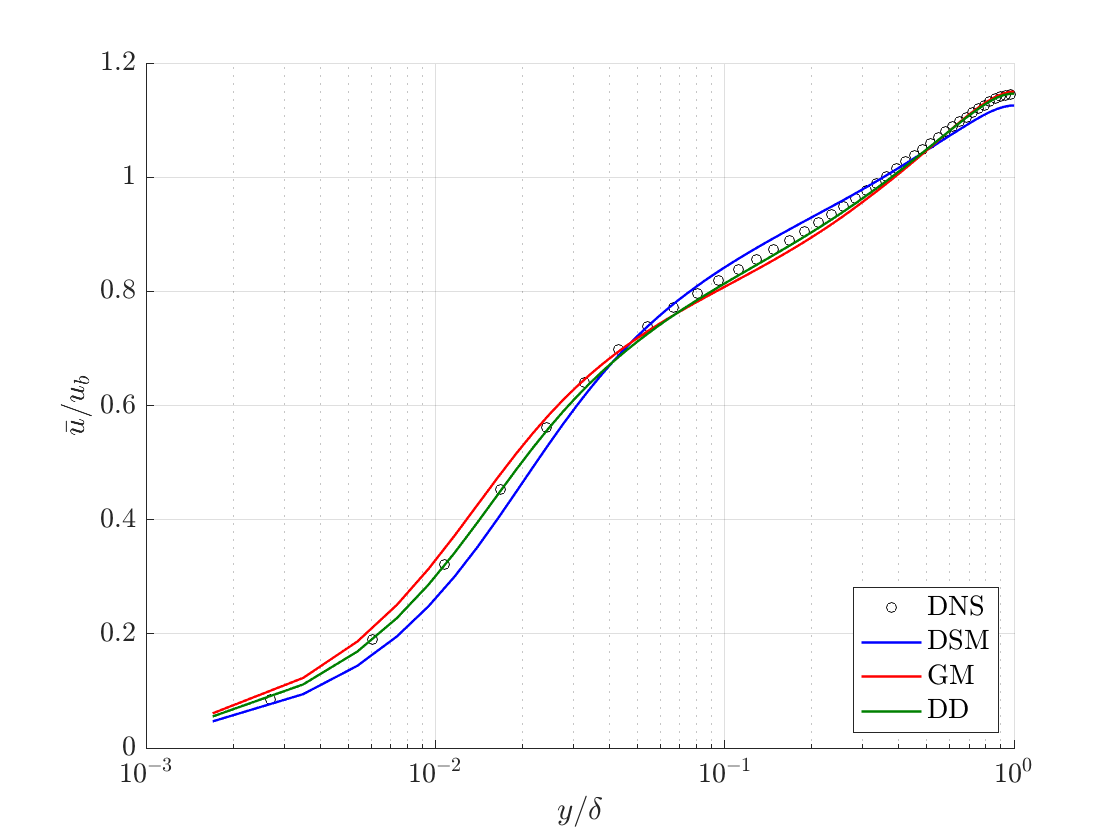}}    
    \caption{Velocity profiles for the a) coarse grid resolution and b) medium grid resolution for turbulent channel flow at $Re_{\tau} = 590$}
    \label{fig:Channel_590_up}
\end{figure}

\begin{figure}[t!]
    \centering
    \subfigure[\label{fig:590_uu_coarse}]{\includegraphics[width=0.49\textwidth]{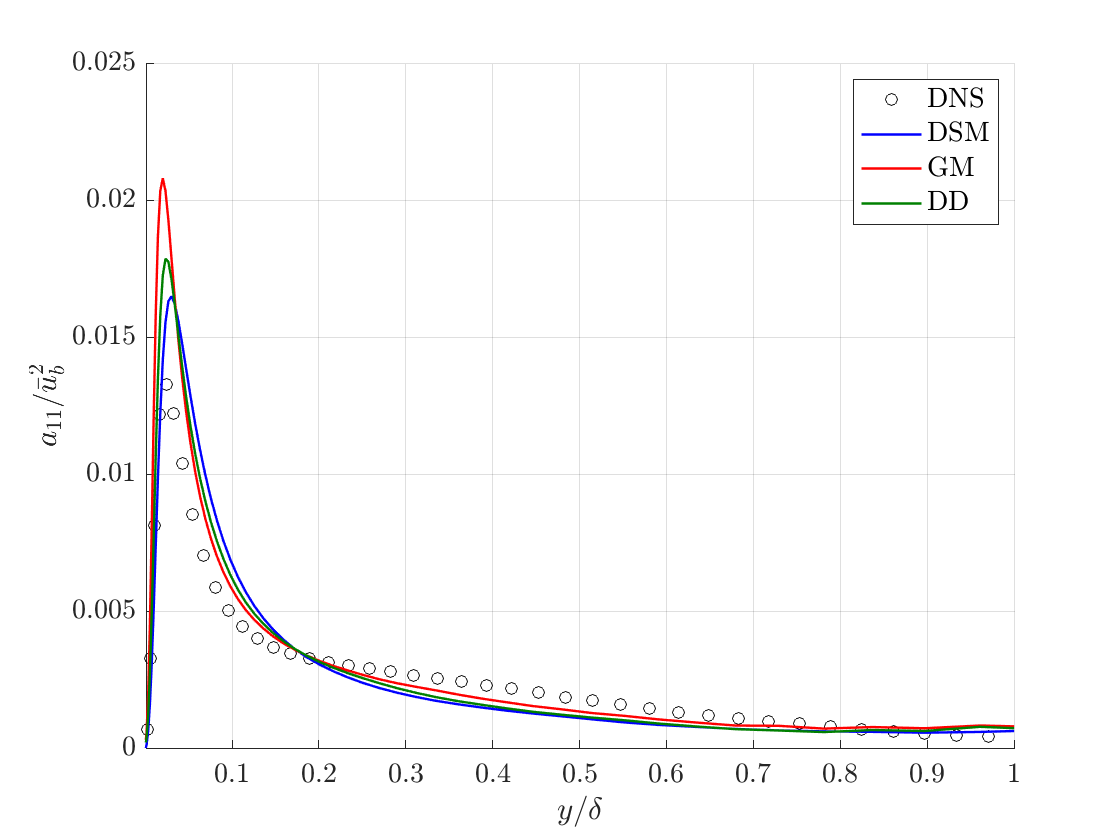}}
    \subfigure[\label{fig:590_vv_coarse}]{\includegraphics[width=0.49\textwidth]{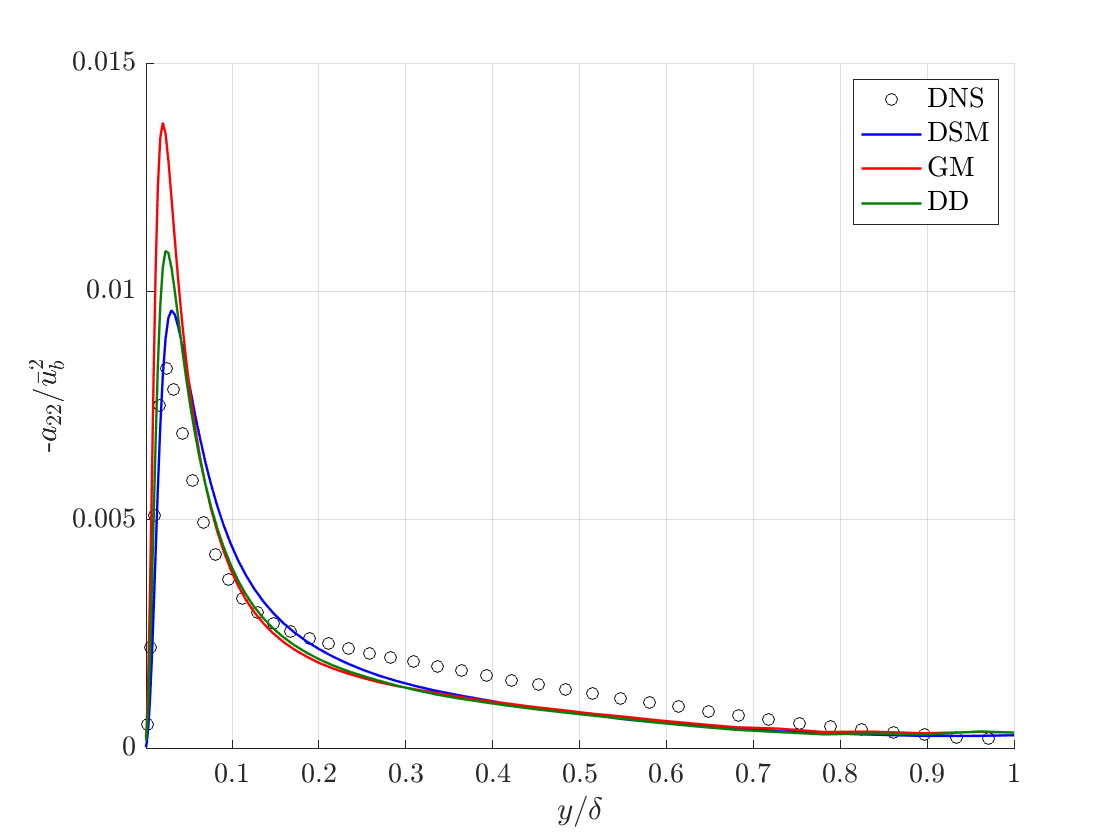}}
    \subfigure[\label{fig:590_ww_coarse}]{\includegraphics[width=0.49\textwidth]{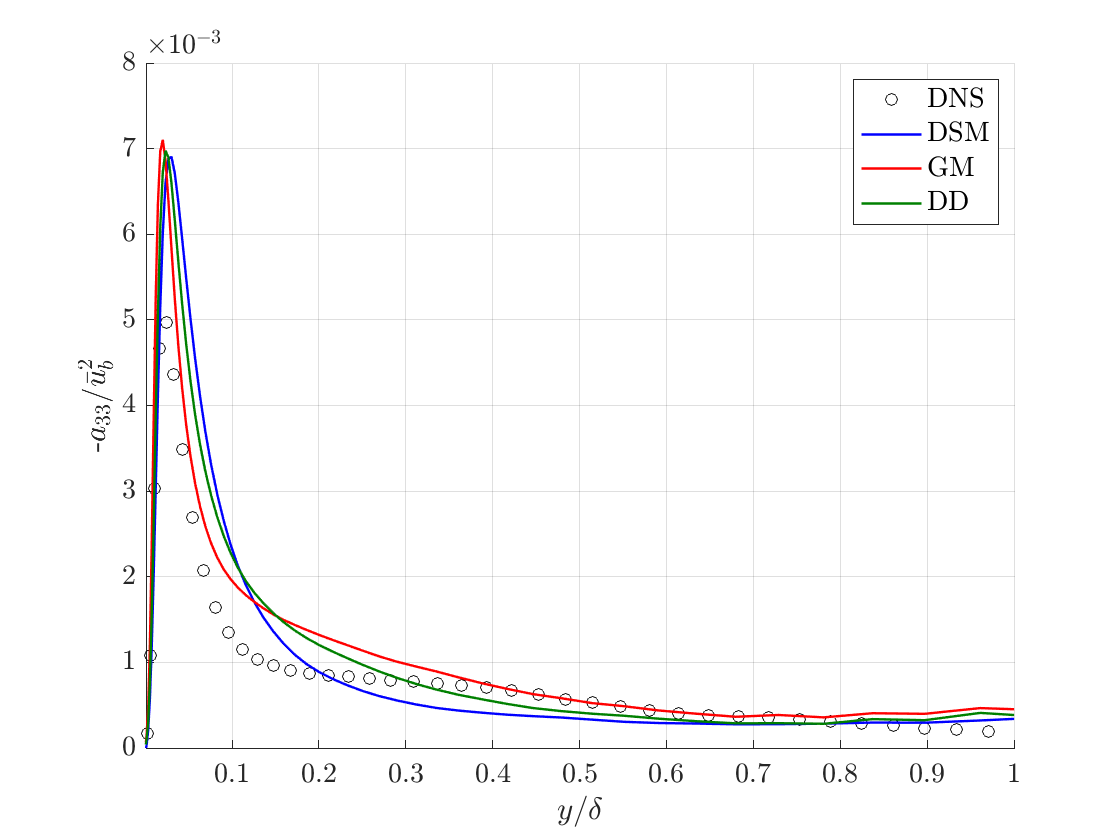}}
    \subfigure[\label{fig:590_uv_coarse}]{\includegraphics[width=0.49\textwidth]{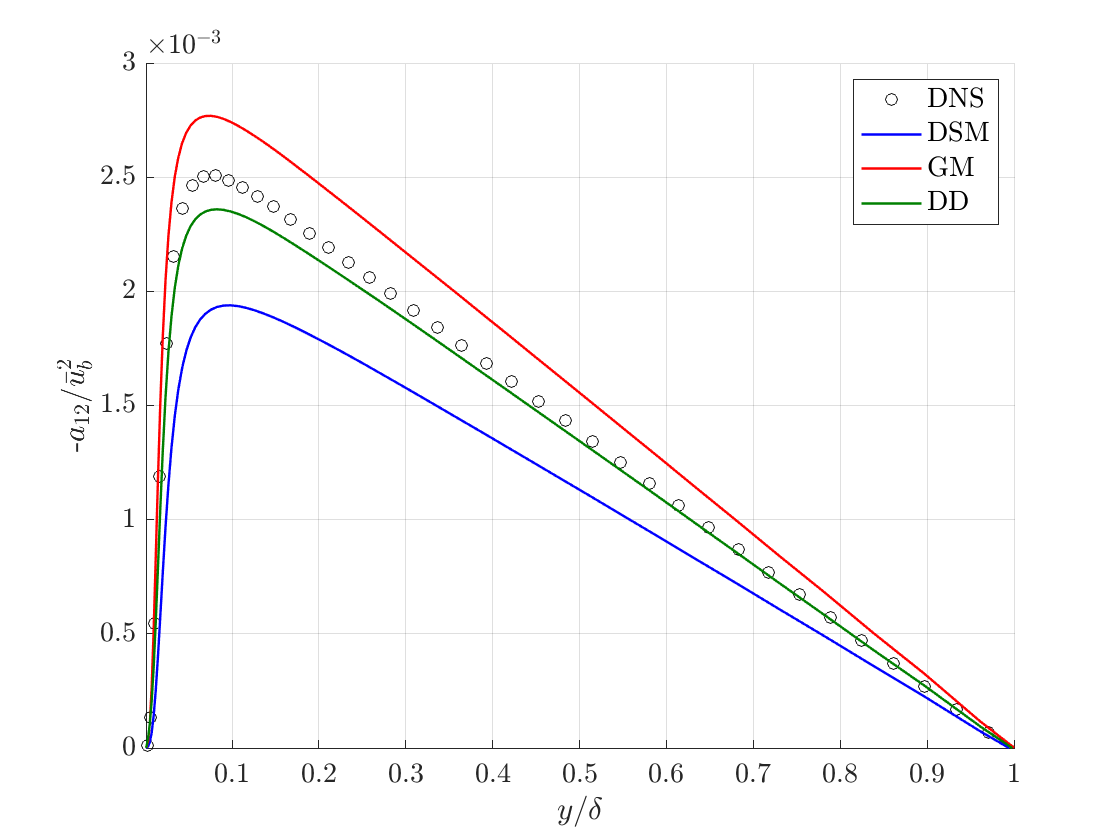}}    
    \caption{ (a) $a_{11}$, (b) $a_{22}$, (c) $a_{33}$ and (d) $a_{12}$ for the coarse grid resolution for turbulent channel flow at $Re_{\tau} = 590$}
    \label{fig:Channel_590_coarse}
\end{figure}

\begin{figure}[t!]
    \centering
    \subfigure[\label{fig:590_uu_fine}]{\includegraphics[width=0.49\textwidth]{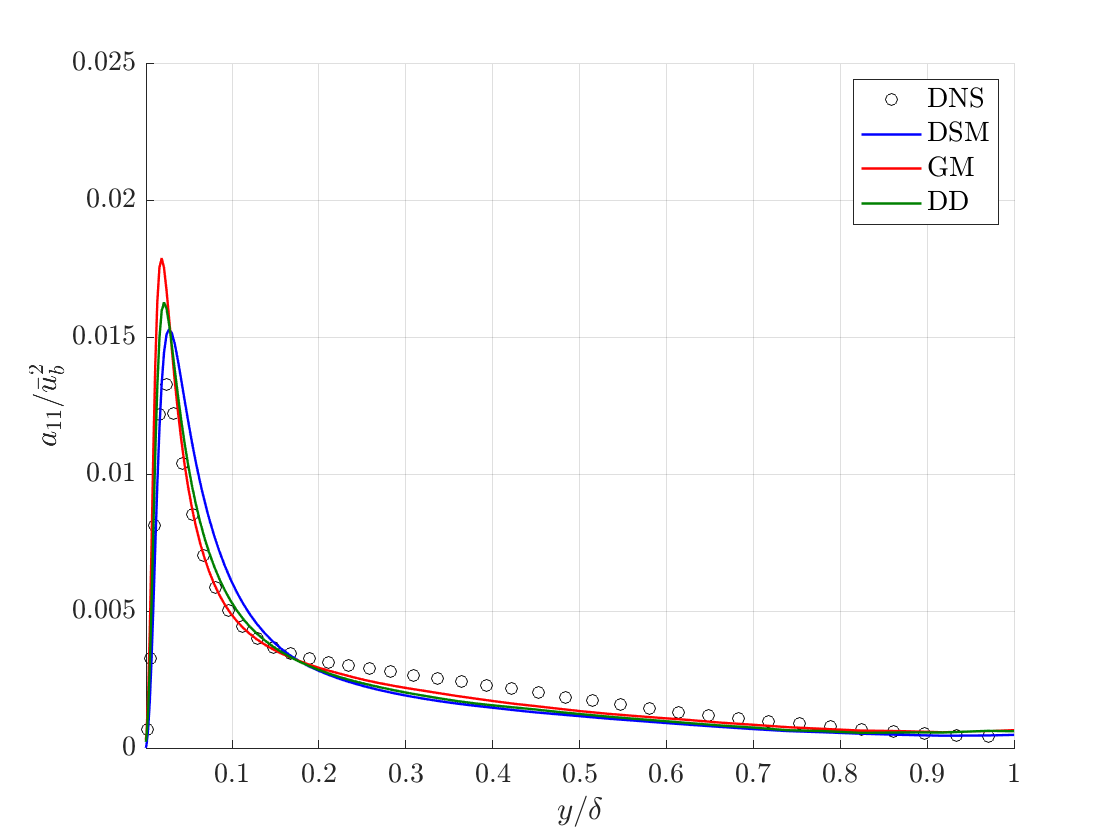}}
    \subfigure[\label{fig:590_vv_fine}]{\includegraphics[width=0.49\textwidth]{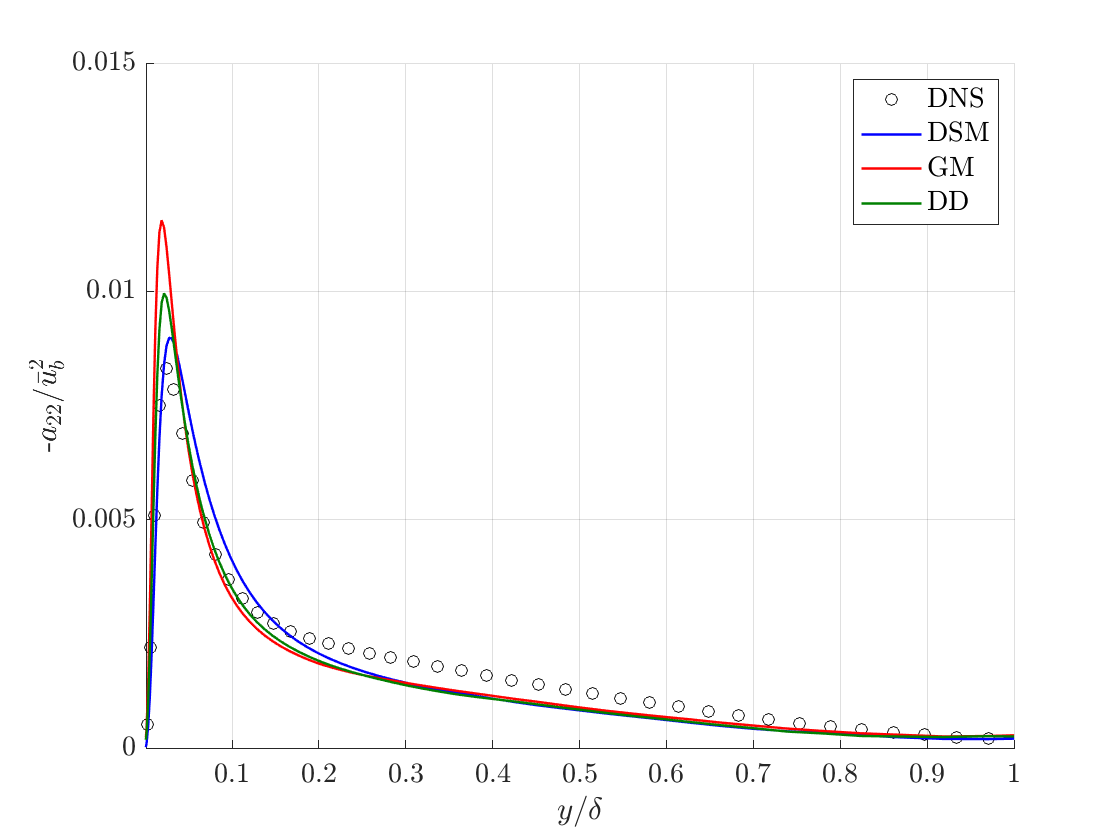}}
    \subfigure[\label{fig:590_ww_fine}]{\includegraphics[width=0.49\textwidth]{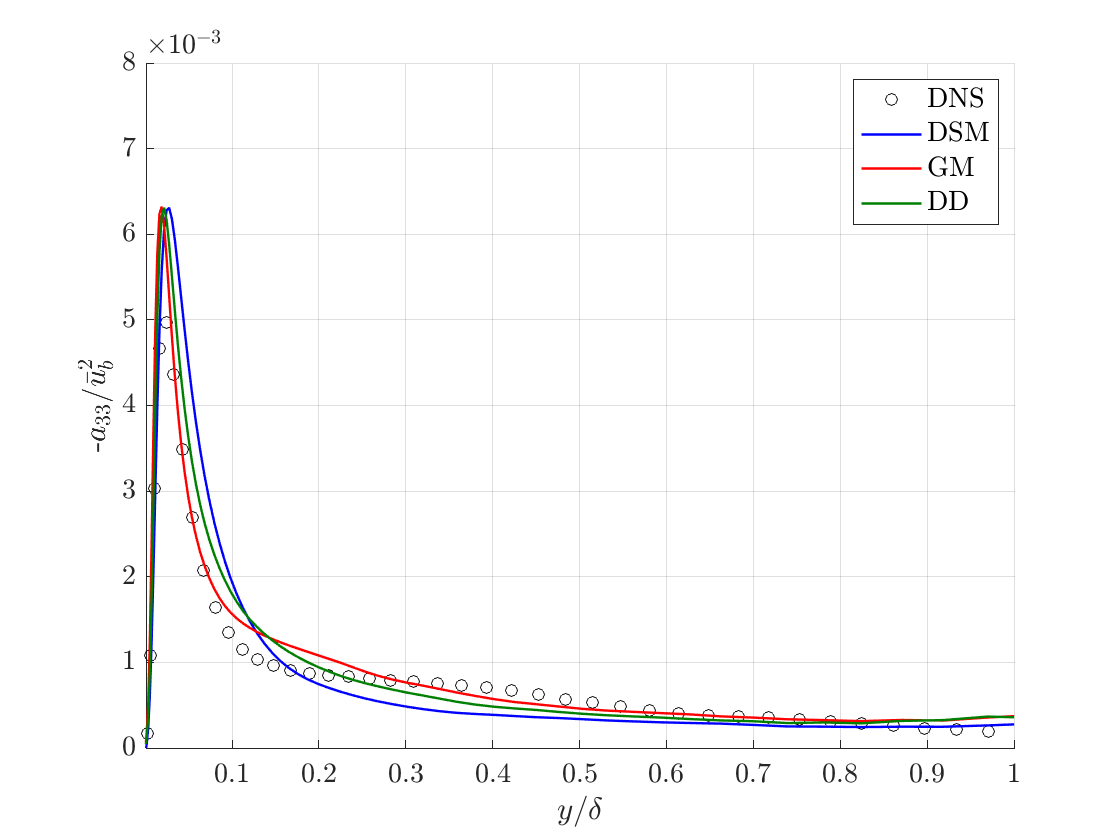}}
    \subfigure[\label{fig:590_uv_fine}]{\includegraphics[width=0.49\textwidth]{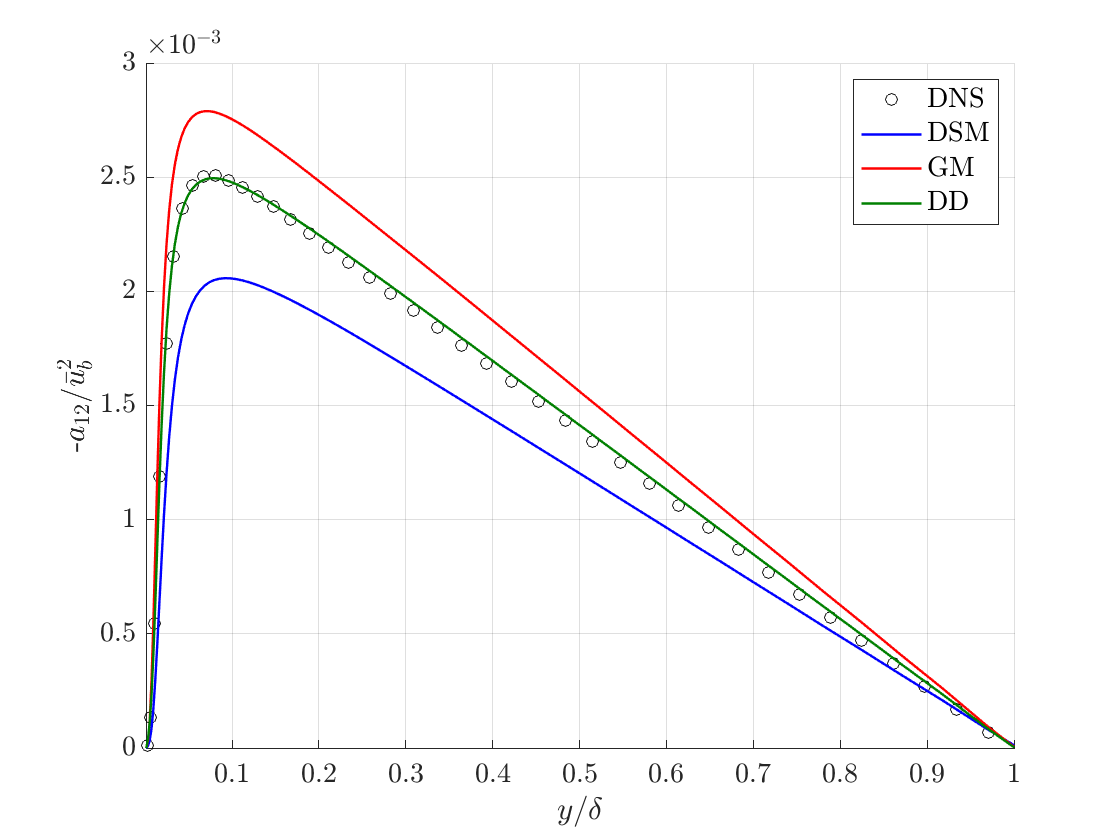}}    
    \caption{ (a) $a_{11}$, (b) $a_{22}$, (c) $a_{33}$ and (d) $a_{12}$ for the medium grid resolution for turbulent channel flow at $Re_{\tau} = 590$}
    \label{fig:Channel_590_fine}
\end{figure}
The mean velocity profile predictions for $Re_{\tau} = 590$ are shown in \figref{Channel_590_up}. We observe that the dynamic Smagorinsky model underpredicts the mean velocity, whereas the gradient model overpredicts the mean velocity profile close to the wall. The data-driven model yields the closest mean velocity profiles to the DNS. The deviatoric part of Reynolds stress tensor predictions for the coarse and fine grid-resolutions for turbulent channel flow at $Re_{\tau} = 590$ are shown in \figref{Channel_590_coarse} and \figref{Channel_590_fine} respectively. All SGS models overpredict the peak Reynolds normal stresses for the coarse grid resolution. The dynamic Smagorinsky model underpredicts the Reynolds shear stress, whereas the gradient model overpredicts the Reynolds shear stress. The data-driven model gives the closest prediction of Reynolds shear stress to the DNS. All SGS models still overpredict the Reynolds normal stresses for the medium grid resolution, although the overprediction is much smaller than the coarse grid resolution case. The data-driven model gives close results to the DNS for Reynolds shear stress, whereas the dynamic Smagorinsky model overpredicts and the gradient model underpredicts Reynolds shear stress significantly.

These turbulent channel flow simulations indicate that the data-driven model yields the closest results to DNS compared to other SGS models considered in this article. Furthermore, these two \textit{a posteriori} test cases highlight that the anisotropic data-driven model predicts more accurate statistics than the existing SGS model for the turbulent flows investigated in this article. Note that both of these \textit{a posteriori} test cases had flow physics outside the training dataset: FHIT was performed at a much higher Reynolds number and flow inside a turbulent channel flow is a wall-bounded shear flow. The good results for both these test cases indicate that the model not only generalizes well for different anisotropy than the training dataset but also to Reynolds number and flow physics outside the training dataset.

\section{Conclusions}
\label{section:Conclusions}

In this article, we proposed an SGS tensor model form applicable for large eddy simulations of turbulent flows using anisotropic grid resolutions. The model form embeds filter anisotropy in addition to physical invariance properties such as Galilean, rotational, reflectional and unit invariance. The filter width anisotropy is embedded in the model form by constructing a mapping from an anisotropic physical space to a parent filter space. This mapping applied to the SGS tensor provides a subgrid stress tensor anisotropy identity which is subsequently used to formulate an isotropic data-driven model in the parent filter space. Furthermore, by considering the gradient of velocity in the parent space as an input, we ensure the Galilean invariance property. Rotational and reflectional invariance is ensured by representing the model outputs and model inputs in the coordinate frame corresponding to the eigenframe of the symmetric part of the gradient of velocity in the parent filter space. Lastly, unit invariance is ensured by applying the Buckingham-Pi theorem. We showed that a first-order Taylor series expansion of the exact SGS stress, equivalent to the anisotropic form of the gradient model, can be exactly represented by the proposed model form. The mapping between model inputs and outputs is learned using neural networks trained using a relatively small amount of anisotropic filtered DNS data from forced HIT flow at $Re_{\lambda} = 418$. The learned data-driven model only requires a single layer of neural network with 20 neurons and exhibits a low model evaluation cost. 

We performed \textit{a priori} and \textit{a posteriori} tests to validate the data-driven model and evaluate its performance outside the training dataset. \textit{A priori} tests involved filter anisotropy of different orientation and aspect ratios than the training set. The data-driven model gave better structural accuracy and dissipative behavior than the anisotropic form of the gradient model. For \textit{a posteriori} tests, we considered forced HIT at $Re_{\lambda} = \infty$ and turbulent channel flow at $Re_{\tau} = 395$ and $Re_{\tau} = 590$. The data-driven model gave the best results for both test cases for several grid resolutions. These tests revealed that the learned data-driven model seems to generalize well for the filter anisotropy tensor, Reynolds numbers and flow physics outside the training dataset. We believe the embedding of filter anisotropy and physical invariance properties have a significant role in the success of the proposed model form. A further improvement in model performance can be achieved by expanding the input space. However, this would require a more complex neural network, thereby increasing the model evaluation cost. Such complex models may be important for accurately predicting complex turbulent boundary layer flows involving smooth body separation \cite{Prakash2023b}. The eventual model selection eventually boils down to a common trade-off between accuracy and cost that comes into play in several fields of computational science and engineering.

\section{Acknowledgements}
The authors would like to acknowledge the Computational and Data-Enabled Science and Engineering (CDS\&E) program of the National Science Foundation (NSF) CBET-1710670, as well as the Transformational Tools and Technologies Project of the National Aeronautics and Space Administration (NASA) 80NSSC18M0147 for funding of this work. Moreover, they thank the Argonne Leadership Computing Facility (ALCF) for the resources on which the simulations and post-processing were performed. 

\bibliographystyle{unsrt}
\bibliography{main.bbl}

\end{document}